\documentclass[12pt,a4paper,reqno]{amsart}
\usepackage[utf8]{inputenc}

\usepackage{amsthm, a4wide}
\usepackage{amsmath}
\usepackage{mathtools}
\usepackage{amssymb}
\usepackage{amsfonts}
\usepackage[table]{xcolor}
\usepackage[linesnumbered,ruled,vlined]{algorithm2e}
\usepackage{indentfirst}
\usepackage{hyperref}
\usepackage{graphicx}
\usepackage{booktabs}
\usepackage{caption,subcaption}
\usepackage{xcolor}
\usepackage{multirow}
\usepackage{multicol}
\usepackage{float}
\usepackage{placeins}
\usepackage{diagbox}
\usepackage{array}
\usepackage{longtable}

\usepackage[numbers,sort&compress]{natbib}
\bibpunct[, ]{[}{]}{,}{n}{,}{,}
\makeatletter
\def\NAT@def@citea{\def\@citea{\NAT@separator}}
\makeatother

\newcolumntype{P}[1]{>{\centering\arraybackslash}p{#1}}

\SetCommentSty{mycommfont}

\SetKwInput{KwInput}{Input}                
\SetKwInput{KwOutput}{Output}  

\newcolumntype{L}[1]{>{\raggedright\arraybackslash}p{#1}}
 
\newcolumntype{C}[1]{>{\centering\arraybackslash}p{#1}}
 
\newcolumntype{R}[1]{>{\raggedleft\arraybackslash}p{#1}}

\usepackage{comment}

\begin{document}





\title[Variable selection in sparse multivariate GLARMA models]{Variable selection in sparse multivariate GLARMA models: Application to germination control by environment}

\author{M. Gomtsyan, C. L\'evy-Leduc, S. Ouadah, L. Sansonnet} 
\address{Université Paris-Saclay, AgroParisTech, INRAE, UMR MIA Paris-Saclay, 91120 Palaiseau, France}
\author{C. Bailly}
\address{UMR7622 CNRS-Sorbonne Université; Laboratoire de Biologie du Développement, Biologie des Semences, Institut de Biologie Paris-Seine, Paris, France.
 Boite 24, 4 place Jussieu, Paris 75005, France}
\author{L. Rajjou}
\address{Institut Jean-Pierre Bourgin, INRAE, AgroParisTech, Université Paris-Saclay, 78026, Versailles, France}

\begin{abstract}
  We propose a novel and efficient iterative two-stage variable selection approach for multivariate sparse GLARMA models, which can be used for modelling multivariate discrete-valued time series.  Our approach consists in iteratively combining two steps: the
estimation of the autoregressive moving average (ARMA) coefficients of multivariate
GLARMA models and the variable selection in the coefficients of the Generalized Linear Model (GLM) part of the model performed by regularized methods.
We explain how to implement our approach efficiently. Then we assess the performance of our methodology using synthetic data and compare it with alternative methods. Finally, we illustrate it on RNA-Seq data resulting from polyribosome profiling to determine translational status for all mRNAs in germinating seeds. Our approach, which is implemented in the \texttt{MultiGlarmaVarSel} R package and available on the CRAN, is very attractive since it benefits from a low computational load and is able to outperform the other methods for recovering the null and non-null coefficients. 
\end{abstract}

\keywords{multivariate GLARMA, sparsity, variable selection, seed quality, gene expression}

\maketitle

\section{Introduction}
\label{seq1}
Seed germination is a complex agronomic trait largely influenced by environmental conditions~\cite{reed:etal:2022}. In cropping systems related to seed production, climatic variations experienced by the mother plant shape the physiological features of seeds, such as dormancy, longevity and germination vigor~\cite{iwasaki:etal:2022}. Plants grown under different temperature regimes produce seeds with contrasting germination potential. The molecular factors that may explain these phenotypes are still poorly described. It has been previously demonstrated that the translation of mRNAs is a key and essential process for the success of germination~\cite{sano:etal:2020}. Studying translation in germinating seeds leads to a better understanding of gene expression regulation providing a direct link between transcriptome and proteome rearrangements~\cite{galland:etal:2014}. Polysome profiling has been developed to infer the translational status of specific mRNA populations~\cite{basbouss:etal:2015, bai:etal:2017}. A rapid polysome formation occurs during early germination process. The combined approaches of polysome profiling and RNA-seq provide a unique opportunity to thoroughly investigate the translational dynamics of germinating seeds produced under different temperature regimes to highlight novel molecular mechanisms related to the physiological quality of seeds in response to the environment of the mother plant.

In this paper we consider a novel multivariate count time series model to study the translational dynamics of germinating seeds. A detailed review of the main approaches for modelling multivariate count time series is available in~\cite{fokianos:2021}. These approaches can be classified into three model classes described hereafter.

The first class includes integer-valued autoregressive (INAR) models.  The first introduction of INAR(1) processes was done by~\cite{McKenzie:1985} and~\cite{al-osh:alzaid:1987}. Later it was extended to $p$th order process in~\cite{al-osh:alzaid:1990}. The properties of the multivariate INAR (MINAR) were derived in~\cite{franke:rao:1995} and~\cite{latour:1997}. Further studies of MINAR were done by~\cite{pedeli:karlis:2013a} and~\cite{pedeli:karlis:2013b}. However, even in the univariate INAR models, the statistical inference is not straightforward, as explained in \cite{davis:etal:2021}, and this is all the more true for higher-order INAR models.

The second class are parameter-driven models. Following the first introduction by~\cite{cox:1981}, parameter-driven models  are time series driven by an unobserved process. It means that the state vector evolves independently of the past history of the observations. Multivariate state space models are studied in~\cite{jorgensen:etal:1996} and \cite{jung:etal:2011}. Additional developments are found in~\cite{ravishanker:etal:2014}. Although these models are simple to construct, the parameter estimation is computationally expensive, see~\cite{jung:2001}.

The third class of models, observation-driven models, do not suffer from computational drawback and are an alternative to parameter-driven models. In these models, the state vector depends on past observations and some additional noise.  Univariate observation-driven models were first proposed by~\cite{cox:1981} and further studied by~\cite{zeger:qaqish:1988}. Different kinds of observation-driven models can be found in the literature: the Generalized Linear Autoregressive Moving Average (GLARMA) models introduced by~\cite{davis:1999} and further studied in~\cite{davis:dunsmuir:streett:2003}, \cite{davis:dunsmuir:street:2005}, \cite{dunsmuir:2015} and  the (log-)linear Poisson autoregressive models studied in~\cite{fokianos:2009}, \cite{fokianos:2011} and~\cite{fokianos:2012}. Note that GLARMA models cannot be seen as a particular case of the log-linear Poisson autoregressive models. In the past years many studies were conducted in the framework of multivariate observation-driven count time series models, many of which are based on the copula approach. An example is the Multivariate Autoregressive Conditional Double Poisson model~\cite{heinen:rengifo:2007}, based on the double Poisson distribution with the mean vector being a VARMA process.
  Another model using copula~\cite{bien:etal:2011}  is developed for count time series with a domain $\mathbb{Z}^n$, $n \in \mathbb{N}$. Here the conditional probabilities of the direction of the process (whether the process is negative, positive or equal to zero) is modeled with the autoregressive conditional multinomial model (ACM). In~\cite{fokianos:etal:2020}, the authors impose a copula function on a vector of related continuous random variables to determine the joint distribution of the count time series. Finally, the model in~\cite{held:hofmann:2005} can be seen as a Poisson branching process model with immigration. It takes as covariates for the mean of each series at time $t$ the counts of other series at time $t-1.$

In our context of analyzing RNA-Seq data from polysome profiling experiments, we are interested in performing variable selection in multivariate count time series. However, this problem is not addressed in the exact framework of our interest so far. There exist methods for variable selection for multivariate Poisson data using spike and slab approach~\cite{giese:etal:2021}. The method is based on extending the Poisson Lognormal model in~\cite{aitchison:1989}, which is a parameter-driven model, to the multivariate case and relaxing the mean-equal-variance property of the Poisson distribution. Another study in~\cite{lee:etal:2018} performs Bayesian variable selection in multivariate zero-inflated count data. 

In this paper, we develop an observation-driven variable selection model, which is an extension of~\cite{gomtsyan:etal:2022} to the multivariate case by considering
the following multivariate GLARMA model. Given the past history $\mathcal{F}_{i,j,t-1} = \sigma(Y_{i,j,s}, s\leq t-1)$, we assume that
\begin{equation}
Y_{i,j,t} | \mathcal{F}_{i,j,t-1} \sim \mathcal{P}({\mu}_{i,j,t}^\star), 
\label{eq1}
\end{equation} 
where $\mathcal{P}(\mu)$ denotes the Poisson distribution with mean $\mu$, $1 \leq i \leq I$, $1 \leq j \leq n_i$ and $1 \leq t \leq T$.
For instance, $Y_{i,j,t}$ can be seen as a random variable modelling RNA-Seq data of the $j$th replication of gene $t$ obtained in condition $i$.
In \eqref{eq1}
\begin{equation}
    \mu_{i,j,t}^\star = \exp (W_{i,j,t}^\star) \quad \text{with} \quad W_{i,j,t}^\star = \eta_{i,t}^\star+ Z_{i,j,t}^\star,
    \label{eq2}
\end{equation}
where
\begin{equation}
Z_{i,j,t}^\star = \sum_{k=1}^q \gamma_k^\star E_{i,j,t}^\star, \quad \quad \text{with } 1 \leq q \leq \infty,
\label{eq3}
\end{equation}
and $\eta_{i,t}^\star$, the non random part of $ W_{i,j,t}^\star$, does not depend on $j$.

Let us denote  $\pmb{\eta}^\star = (\eta_{1,1}^\star, \dots, \eta_{I,1}^\star,  \eta_{I,2}^\star, \dots, \eta_{I,T}^\star)'$ the vector of 
coefficients corresponding to the effect of a qualitative variable on the observations.
For instance, in the case of RNA-Seq data, $\eta_{i,t}^\star$ can be seen as the effect of condition $i$ (i.e. temperature regime during seed production) on  polysome-associated mRNAs $t$.
Assume moreover that $\pmb{\gamma}^\star =  (\gamma_{1}^\star , \dots, \gamma_{q}^\star )'$ is such that $\sum_{k\geq 1}|\gamma_k^\star|<\infty$, where
$u'$ denotes the transpose of $u$. Additionally, 
\begin{equation}
E_{i,j,t}^\star = \frac{Y_{i,j,t} - \mu_{i,j,t}^\star}{\mu_{i,j,t}^\star} = Y_{i,j,t} \exp \big(-W_{i,j,t}^\star \big) - 1.
\label{eq4}
\end{equation}
with $E_{i,j,t}^\star = 0$ for all $t \leq 0$ and $1 \leq q \leq \infty$. When $q=\infty$,  $Z_{i,j,t}^\star$ satisfies an ARMA-like recursion in \eqref{eq4},
because causal ARMA can be written as MA process of infinite order.

$E_{i,j,t}^\star$ in \eqref{eq4} corresponds to the particular case of working residuals in classical Generalized Linear Models (GLM) usually defined by $E_{i,j,t}^\star=(Y_{i,j,t}- \mu_{i,j,t}^\star){\mu_{i,j,t}^\star}^{-\lambda}$ with $\lambda=1$.
The resulting model defined by Equations \eqref{eq1}, \eqref{eq2}, \eqref{eq3} and \eqref{eq4} is referred to as multivariate GLARMA model.

The main goal of this paper is to introduce a novel variable selection approach in the deterministic part ($\pmb{\eta}^\star$) of the
sparse multivariate GLARMA model that is
defined in Equations \eqref{eq1}, \eqref{eq2}, \eqref{eq3} and \eqref{eq4}, where the vector of the $\eta_{i,t}^\star$'s is sparse. Sparsity means that many $\eta_{i,t}^\star$'s are null, and thus just a few coefficients are significant. The novel approach that we propose combines a procedure for estimating the ARMA part coefficients to take into account the dependence that may exist in the data with regularized methods designed for GLM as those proposed by \cite{friedman:hastie:tibshirani:2010} and \cite{hastie2019statistical}.

The paper is organized as follows. Firstly, we propose a novel two-stage estimation procedure in multivariate GLARMA models in Section \ref{sec2.1} and Section \ref{sec2.2}. It consists of first estimating the ARMA coefficients and then estimating the $\eta_{i,t}^\star$'s by using a regularized approach. The practical implementation of our approach is given in Section \ref{sec2.3}.
Next, in Section \ref{sec3}, we provide numerical experiments to illustrate our method and compare its performance to alternative approaches on finite sample size data. Additionally, in Section \ref{sec4}, we apply our method to RNA-Seq data from polysome profiling experiments to determine translational status for all mRNAs in germinating seeds.

\section{Statistical Inference}
\label{sec2}

Extending the estimation procedure existing in standard univariate GLARMA models described in \cite{davis:dunsmuir:streett:2003} and
\cite{davis:dunsmuir:street:2005} to the multivariate case
would consist in estimating
$\pmb{\delta}^\star = (\pmb{\eta}^{\star\prime}, \pmb{\gamma}^{\star\prime})$, where $\pmb{\eta}^{\star\prime}$ is the vector of coefficients and $\pmb{\gamma}^{\star\prime}$ is the vector of the ARMA part coefficients by $\pmb{\widehat{\delta}}$, which is defined as follows:
\begin{equation}
\pmb{\widehat{\delta}} = \underset{\pmb{\delta}}{\arg\max} \, L(\pmb{\delta}).
\label{eq7}
\end{equation}
In \eqref{eq7}, $L$ is based on the conditional log-likelihood and is defined by: 
\begin{equation*}
L(\pmb{\delta}) = \sum_{i=1}^I \sum_{j=1}^{n_i} \sum_{t=1}^T (Y_{i,j,t} W_{i,j,t}(\pmb{\delta})- \exp (W_{i,j,t} (\pmb{\delta})),
\end{equation*}
where $W_{i,j,t}(\pmb{\delta})$ is defined as in \eqref{eq2}-\eqref{eq4}:
\begin{equation}\label{eq:Wijt}
W_{i,j,t}(\pmb{\delta})= \eta_{i,t} + \sum_{k=1}^q \gamma_k E_{i,j,t}(\pmb{\delta}) \textrm{ with }
E_{i,j,t}(\pmb{\delta})=Y_{i,j,t} \exp \big(-W_{i,j,t}(\pmb{\delta}) \big) - 1.
\end{equation}

However, this procedure is not designed for dealing with a sparse framework where many components of $\pmb{\eta}^\star$ are null.
This is the reason why we propose hereafter a novel two-stage estimation procedure described in the following sections.


\subsection{Estimation of $\pmb{\gamma^{\star}}$}
\label{sec2.1}

In our estimation procedure, we use the Newton-Raphson algorithm to obtain $\pmb{\widehat{\gamma}}$ based on the following recursion. For $r \geq 1$, starting from the initial value $\pmb{\gamma}^{(0)}= (\gamma_1^{(0)}, \dots, \gamma_q^{(0)})^{'}$ and $\pmb{\eta}^{(0)} = (\eta_{1,1}^{(0)}, \dots, \eta_{I,1}^{(0)}, \eta_{I,2}^{(0)}, \dots, \eta_{I,T}^{(0)})^{'}$:
\begin{equation}
    \pmb{\gamma}^{(r)} = \pmb{\gamma}^{(r-1)} - \frac{\partial^2 L}{\partial \pmb{\gamma}^{'} \partial \pmb{\gamma}} \big(\pmb{\eta}^{(0)}, \pmb{\gamma}^{(r-1)}\big)^{-1} \frac{\partial L}{\partial \pmb{\gamma}} \big(\pmb{\eta}^{(0)}, \pmb{\gamma}^{(r-1)} \big).
    \label{eq10}
\end{equation}

To obtain $\frac{\partial L}{\partial \pmb{\gamma}}$, we shall use that
\begin{equation*}
  \frac{\partial L}{\partial \pmb{\gamma}}(\pmb{\eta}^{(0)}, \pmb{\gamma}) = \sum_{i=1}^I \sum_{j=1}^{n_i} \sum_{t=1}^T (Y_{i,j,t} - \exp (W_{i,j,t} (\pmb{\eta}^{(0)}, \pmb{\gamma})))
  \frac{\partial W_{i,j,t}(\pmb{\eta}^{(0)}, \pmb{\gamma})}{\partial \pmb{\gamma}},
\end{equation*}
where details for computing the first derivative of $W_{i,j,t}(\pmb{\eta}^{(0)}, \pmb{\gamma})$ with respect to $\pmb{\gamma}$ are given
in Appendix \ref{app1}.

Concerning the Hessian of $L$, it can be obtained as follows:
\begin{align*}
  \frac{\partial^2 L}{\partial \pmb{\gamma}^{'} \partial \pmb{\gamma}}(\pmb{\eta}^{(0)},\pmb{\gamma})
  &= \sum_{i=1}^I \sum_{j=1}^{n_i} \sum_{t=1}^T (Y_{i,j,t} - \exp (W_{i,j,t} (\pmb{\eta}^{(0)},\pmb{\gamma}))) \frac{\partial^2 W_{i,j,t}(\pmb{\eta}^{(0)},\pmb{\gamma})}{\partial \pmb{\gamma}' \partial \pmb{\gamma}} \\ &- \sum_{i=1}^I \sum_{j=1}^{n_i} \sum_{t=1}^T \exp (W_{i,j,t} (\pmb{\eta}^{(0)},\pmb{\gamma}))
                                                                                                                                                                     \frac{\partial W_{i,j,t}(\pmb{\eta}^{(0)},\pmb{\gamma})}{\partial \pmb{\gamma}'} \frac{\partial W_{i,j,t}(\pmb{\eta}^{(0)},\pmb{\gamma})}{\partial \pmb{\gamma}},
\end{align*}
where details for computing the second derivative of $W_{i,j,t}(\pmb{\eta}^{(0)}, \pmb{\gamma})$ with respect to $\pmb{\gamma}$ are given
in Appendix \ref{app2}.

\subsection{Variable selection in $\pmb{\eta^{\star}}$ estimation}
\label{sec2.2}

\subsubsection{Variable selection criterion}
To perform variable selection in the $\pmb{\eta^{\star}}$ of Model \eqref{eq2}--(\ref{eq4}), namely to obtain a sparse estimator of $\pmb{\eta^{\star}}$, we shall use a regularized approach inspired by \cite{friedman:hastie:tibshirani:2010}  for fitting sparse generalized linear models. It consists in penalizing
  (with an $\ell_1$ penalty) a quadratic approximation to the log-likelihood obtained by a second order Taylor expansion. Using $\pmb{\eta}^{(0)}$ and $\widehat{\pmb{\gamma}}$ defined in Section \ref{sec2.1}, we obtain the quadratic approximation as follows:
\begin{equation*}
\begin{split}
\Tilde{L}(\pmb{\eta}) & \coloneqq  L(\eta_{1,1}, \dots, \eta_{I,1}, \eta_{I,2}, \dots, \eta_{I,T}, \widehat{\pmb{\gamma}})\\
 & = \Tilde{L}(\pmb{\eta}^{(0)}) + \frac{\partial L}{\partial \pmb{\eta}}(\pmb{\eta}^{(0)}, \widehat{\pmb{\gamma}})(\pmb{\eta}-\pmb{\eta}^{(0)}) \\
 & \qquad \qquad +  \frac{1}{2}(\pmb{\eta}-\pmb{\eta}^{(0)})' \frac{\partial^2 L}{\partial \pmb{\eta}\partial \pmb{\eta}^{'}}(\pmb{\eta}^{(0)}, \widehat{\pmb{\gamma}})(\pmb{\eta}-\pmb{\eta}^{(0)}),
\end{split}
\end{equation*}
where 
\begin{equation*}
\frac{\partial L}{\partial \pmb{\eta}} = \Big(\frac{\partial L}{\partial \eta_{1,1}}, \dots, \frac{\partial L}{\partial \eta_{I,1}}, \frac{\partial L}{\partial \eta_{I,2}}, \dots, \frac{\partial L}{\partial \eta_{I,T}} \Big)
\end{equation*}
and
\begin{equation*}
\frac{\partial^2 L}{\partial \pmb{\eta}\partial{\pmb{\eta}}^{'}} = \Big( \frac{\partial^2 L }{\partial \eta_{i_0,t_0} \partial \eta_{i_1,t_1}}\Big)_{\substack{1 \leq i_0, i_1 \leq I\\1 \leq t_0, t_1 \leq T}}.
\end{equation*}

Let $U\Lambda U'$ be the singular values decomposition of the positive semidefinite symmetric matrix $-\frac{\partial^2 L}{\partial \pmb{\eta}\pmb{\eta}^{'}}(\pmb{\eta}^{(0)} ,\widehat{\pmb{\gamma}})$ and $\pmb{\nu} - \pmb{\nu}^{(0)} = U'(\pmb{\eta} - \pmb{\eta}^{(0)})$. Therefore, the quadratic approximation is

\begin{equation}\label{eq:Tilde_L}
    \Tilde{L}(\pmb{\eta}) =  \Tilde{L}(\pmb{\eta}^{(0)}) + \frac{\partial L}{\partial \pmb{\eta}}(\pmb{\eta}^{(0)}, \widehat{\pmb{\gamma}}) U (\pmb{\nu} - \pmb{\nu}^{(0)}) - \frac{1}{2}(\pmb{\nu} - \pmb{\nu}^{(0)})' \Lambda (\pmb{\nu} - \pmb{\nu}^{(0)}).
\end{equation}

In order to obtain a sparse estimator of $\pmb{\eta}^{\star}$ we use $\widehat{\pmb{\eta}}(\lambda)$ defined by minimizing the following criterion:
\begin{equation}
    \widehat{\pmb{\eta}}(\lambda) = \underset{\pmb{\eta}}{\arg \min} \{ -\Tilde{L}_Q(\pmb{\eta}) + \lambda \|\pmb{\eta}\|_1\},
    \label{eq16}
\end{equation}
for a positive $\lambda$, where $\|\eta\|_1 = \sum_{i=1}^I \sum_{t=1}^T |\eta_{i,t}|$ and $\Tilde{L}_Q(\pmb{\eta})$ denotes the quadratic approximation of the log-likelihood. This quadratic approximation is defined by
\begin{equation}\label{eq:LQ}
-\Tilde{L}_Q(\pmb{\eta}) =  \frac{1}{2} \|\mathcal{Y}- \mathcal{X} \pmb{\eta}\|_2^2
\end{equation}
with 
\begin{equation}
    \mathcal{Y} = \Lambda^{1/2} U' \pmb{\eta}^{(0)} + \Lambda^{-1/2} U' \Big(\frac{\partial L}{\partial \pmb{\eta}} (\pmb{\eta}^{(0)}, \widehat{\pmb{\gamma}})\Big)', \quad \mathcal{X} = \Lambda^{1/2} U',
    \label{eq18}
\end{equation}
where $\|\cdot\|_2$ is the $\ell_2$ norm.

\subsubsection{Criterion derivation}

Let us now explain how the expression of $\tilde{L}_Q$ given in (\ref{eq:LQ}) was obtained.
By (\ref{eq:Tilde_L}), we get
\begin{equation} 
\begin{split}
\Tilde{L}(\pmb{\eta}) & =  \Tilde{L}(\pmb{\eta}^{(0)}) + \sum_{i=1}^I \sum_{t=1}^T \Bigg( \frac{\partial L}{\partial \pmb{\eta}} (\pmb{\eta}^{(0)}, \widehat{\pmb{\gamma}}) U \Bigg)_{i,t} (\nu_{i,t} - \nu_{i,t}^{(0)}) -\frac{1}{2} \sum_{i=1}^I \sum_{t=1}^T \lambda_{i,t} (\nu_{i,t} - \nu_{i,t}^{(0)})^2 \\
 & = \Tilde{L}(\pmb{\eta}^{(0)}) - \frac{1}{2} \sum_{i=1}^I \sum_{t=1}^T \lambda_{i,t} \Bigg( \nu_{i,t} - \nu_{i,t}^{(0)} - \frac{1}{\lambda_{i,t}} \Bigg( \frac{\partial L}{\partial \pmb{\eta}} (\pmb{\eta}^{(0)}, \widehat{\pmb{\gamma}}) U \Bigg)_{i,t} \Bigg)^2 \\ 
 & \qquad \qquad \quad + \sum_{i=1}^I \sum_{t=1}^T \frac{1}{2 \lambda_{i,t}} \Bigg( \frac{\partial L}{\partial \pmb{\eta}} (\pmb{\eta}^{(0)}, \widehat{\pmb{\gamma}}) U \Bigg)_{i,t}^2,
 \label{eq19}
 \end{split}
\end{equation}
where the $\lambda_{i,t}$'s are the diagonal terms of $\Lambda$.

Since only the second term of \eqref{eq19} depends on $\pmb{\eta}$,
\begin{equation*}
\begin{split}
-\Tilde{L}_Q(\pmb{\eta}) & = \frac{1}{2} \sum_{i=1}^I \sum_{t=1}^T \lambda_{i,t} \Bigg( \nu_{i,t} - \nu_{i,t}^{(0)}  - \frac{1}{\lambda_{i,t}} \Bigg( \frac{\partial L}{\partial \pmb{\eta}} (\pmb{\eta}^{(0)}, \widehat{\pmb{\gamma}})U \Bigg)_{i,t} \Bigg)^2   \\
 & = \frac{1}{2} \Bigg\| \Lambda^{1/2} \Bigg( \pmb{\nu} - \pmb{\nu}^{(0)} - \Lambda^{-1} \Bigg( \frac{\partial L}{\partial \pmb{\eta} } (\pmb{\eta}^{(0)}, \widehat{\pmb{\gamma}}) U \Bigg)' \Bigg) \Bigg\| _2^2 \\ 
 & =  \frac{1}{2} \Bigg\| \Lambda^{1/2} U' ( \pmb{\eta} - \pmb{\eta}^{(0)})  - \Lambda^{-1/2} U' \Bigg( \frac{\partial L}{\partial \pmb{\eta} } (\pmb{\eta}^{(0)}, \widehat{\pmb{\gamma}}) \Bigg)'  \Bigg\| _2^2 \\ 
  & =  \frac{1}{2} \Bigg\| \Lambda^{1/2} U' ( \pmb{\eta}^{(0)} - \pmb{\eta})  + \Lambda^{-1/2} U' \Bigg( \frac{\partial L}{\partial \pmb{\eta} } (\pmb{\eta}^{(0)}, \widehat{\pmb{\gamma}}) \Bigg)'  \Bigg\| _2^2 \\ 
  & = \frac{1}{2} \|\mathcal{Y}- \mathcal{X} \pmb{\eta}\|_2^2,
\end{split} 
\end{equation*}
where
\begin{equation*}
    \mathcal{Y} = \Lambda^{1/2} U' \pmb{\eta}^{(0)} + \Lambda^{-1/2} U' \Big(\frac{\partial L}{\partial \pmb{\eta}} (\pmb{\eta}^{(0)}, \widehat{\pmb{\gamma}})\Big)', \quad \mathcal{X} = \Lambda^{1/2} U'.
\end{equation*}

\subsubsection{Stability selection}\label{subsec:stab_sel}
To obtain the final estimator $\widehat{\pmb{\eta}}$ of $\pmb{\eta^{\star}}$, we shall consider an approach called stability selection devised by \cite{meinshausen:buhlmann:2010}, which guarantees the robustness of the selected variables. This  approach can be described as follows.
The vector $\mathcal{Y}$ defined in (\ref{eq18}) is randomly split into several subsamples of size $IT/2$, corresponding to half of the length of $\mathcal{Y}$. The number of subsamples is equal to 1000 in our numerical experiments.
For each subsample $\mathcal{Y}^{(s)}$ and the corresponding design matrix $\mathcal{X}^{(s)}$,
Criterion \eqref{eq16} is applied with a given $\lambda$,
where $\mathcal{Y}$ and $\mathcal{X}$ are replaced by $\mathcal{Y}^{(s)}$ and $\mathcal{X}^{(s)}$, respectively.
For each subsampling, the indices $i$ and $t$ of the non-null $\widehat{\eta}_{i,t}$ are stored.
In the end, we calculate the frequency of index selection, namely the number of times each couple of indices was selected divided by the number of subsamples. For a given threshold, in the final set of selected variables, we keep the ones whose indices have a frequency larger than this threshold. Concerning the choice of $\lambda$, we shall consider  the smallest element of the grid of $\lambda$ provided by the R \texttt{glmnet} package.
It is also possible to use the one obtained by cross-validation (Chapter 7 of \cite{hastie2009elements}). However, based on our experiments,
choosing the minimal $\lambda$ of the grid led to better results.

\subsection{Practical implementation}
\label{sec2.3}

In practice, the previous approach can be summarized as follows.
\begin{itemize}
\item\textsf{Initialization.} We take for $\boldsymbol{\eta}^{(0)}$  the estimator of $\boldsymbol{\eta}^\star$ obtained by fitting a GLM to the observations $Y_{1,1,1},\dots,Y_{I, n_{I}, T}$ thus ignoring the ARMA part of the model. For $\boldsymbol{\gamma}^{(0)}$, we take the null vector.
\item\textsf{Newton-Raphson algorithm.} We use the recursion defined in \eqref{eq10} with
the initialization $(\boldsymbol{\eta}^{(0)},\boldsymbol{\gamma}^{(0)})$ obtained in the previous step and
we stop at the iteration $R$ such that $\|\boldsymbol{\gamma}^{(R)}-\boldsymbol{\gamma}^{(R-1)}\|_\infty<10^{-6}$.
\item\textsf{Variable selection.} To obtain a sparse estimator of $\boldsymbol{\eta}^\star$, we use Criterion \eqref{eq16},
  where $\boldsymbol{\eta}^{(0)}$ and $\widehat{\boldsymbol{\gamma}}$ appearing in \eqref{eq18} are replaced by $\boldsymbol{\eta}^{(0)}$ and
  $\boldsymbol{\gamma}^{(R)}$ 
obtained in the previous steps. We thus get $\widehat{\boldsymbol{\eta}}$ by using the stability selection approach described in Section \ref{subsec:stab_sel}.

\end{itemize}

This procedure can be improved by iterating the \textsf{Newton-Raphson algorithm} and \textsf{Variable selection} steps. More precisely, let us denote by $\boldsymbol{\eta}_{1}^{(0)}$, $\gamma_{1}^{(R_1)}$ and $\widehat{\boldsymbol{\eta}}_1$ 
the values of $\boldsymbol{\eta}^{(0)}$, $\gamma^{(R)}$ and $\widehat{\boldsymbol{\eta}}$ obtained
in the three steps described above at the first iteration. At the second iteration, $(\boldsymbol{\eta}^{(0)},\boldsymbol{\gamma}^{(0)})$ appearing in the \textsf{Newton-Raphson algorithm} step is replaced by $(\widehat{\boldsymbol{\eta}}_1,\gamma_{1}^{(R_1)})$. At the end of this second iteration, $\widehat{\boldsymbol{\eta}}_2$ and $\gamma_{2}^{(R_2)}$ denote the obtained values of $\widehat{\boldsymbol{\eta}}$ and $\gamma^{(R)}$, respectively. This approach is iterated until the
stabilisation of $\gamma_{k}^{(R_k)}$.

\section{Numerical experiments}
\label{sec3}

This section aims at investigating the performance of our method, which is implemented in the R package \texttt{MultiGlarmaVarSel} available on the
CRAN (Comprehensive R Archive Network). We study it both from a statistical and a numerical point of view, using synthetic data generated from the model defined by \eqref{eq1}--\eqref{eq4}, where $n_i=J$ for all $i$. The different simulation settings that we considered are given in Table \ref{tab1}. In all the experiments we set the number of non-null coefficients in $\pmb{\eta^{\star}}$ to $10$ and the number of simulations to $50$. The non-null values of $\pmb{\eta^{\star}}$
range from 0.41 to 2.62.

\begin{table}[H]
\centering
\begin{tabular}{|c|c|c|c|c|c|c|c|}
\hline
$T$ & $J$ & $I$ & $q^{\star}$ & $\boldsymbol{\gamma^{\star}}$  \\ \hline
50	& 10		& 3  & 1          & 0.5		\\ \hline
50	& 100	& 3	& 1          & 0.5           \\ \hline
200	& 10		& 3 	& 1          & 0.5         	  \\ \hline
200	& 100	& 3   & 1          & 0.5             \\ \hline
50	& 10		& 3   & 2          & 0.2, 0.5       \\ \hline
50	& 100	& 3   & 2          & 0.2, 0.5         \\ \hline
200	& 10		& 3   & 2          & 0.2, 0.5        \\ \hline
200	& 100	& 3   & 2          & 0.2, 0.5        \\ \hline
\end{tabular}
\caption{Parameters of simulated datasets used in the experiments.}
\label{tab1}
\end{table}
\subsection{Statistical performance}
\subsubsection{Estimation of $\pmb{\eta^{\star}}$}
\paragraph{Support recovery of $\pmb{\eta^{\star}}$}
\label{seq3.1}
In this section, we assess the performance of our methodology in terms of support recovery, namely the identification of the
non-null coefficients of $\boldsymbol{\eta}^\star$, and of the estimation of $\boldsymbol{\gamma}^\star$.

\begin{figure}[!htb]
\centering
\includegraphics[scale=0.25]{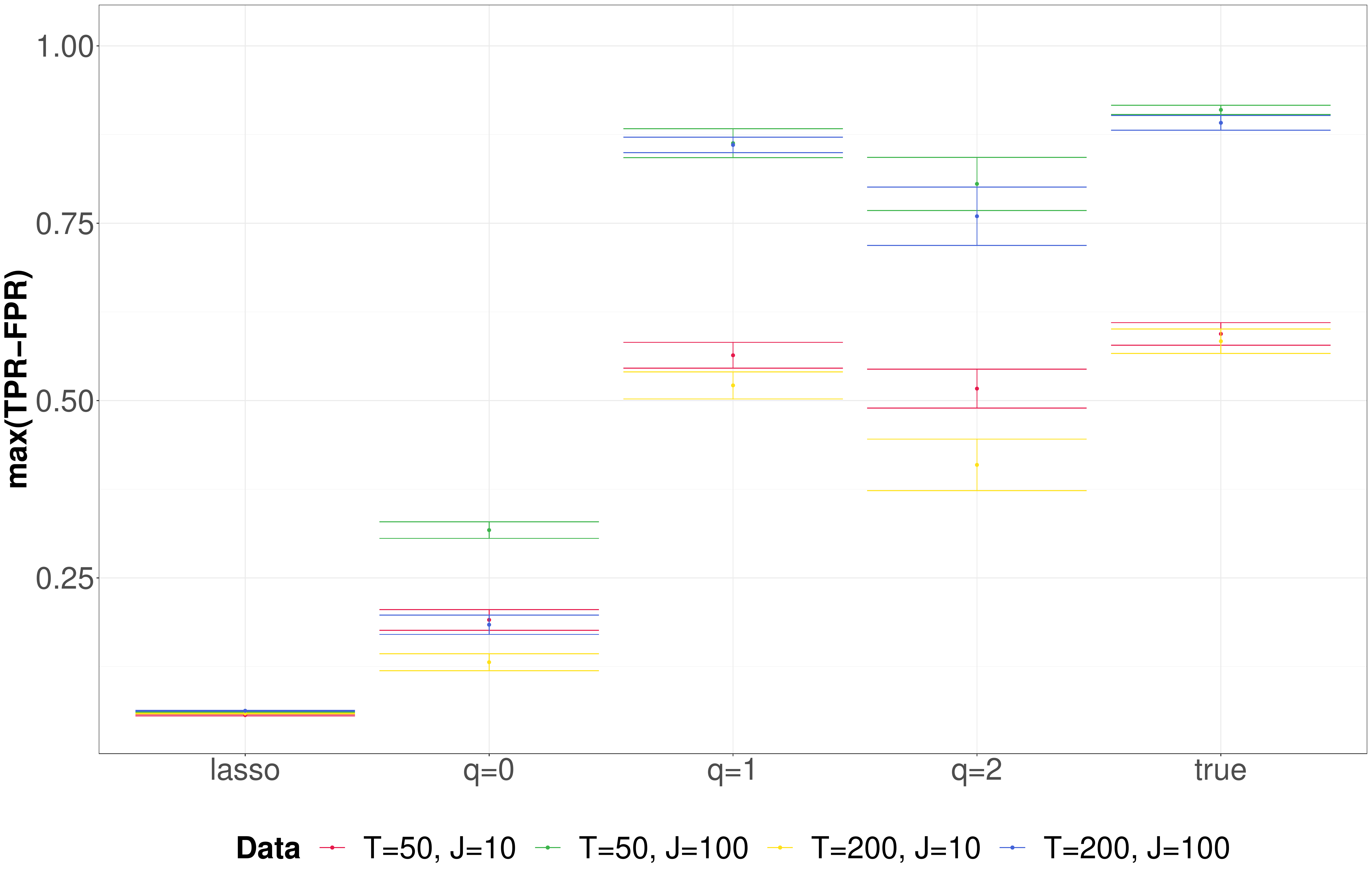}
\caption{Error bars of the maximum difference between TPR and FPR for different thresholds associated to the support recovery of $\boldsymbol{\eta}^\star$
  with 5 approaches for 4 simulation frameworks when $I=3$, $q^{\star}$ = 1, 10 non-null coefficients in $\pmb{\eta^{\star}}$, and 50 simulations.}
\label{fig1}
\end{figure}

Figures \ref{fig1} and \ref{fig3} display the maximum difference between TPR (True Positive Rates, namely the proportion of non-null coefficients correctly estimated as non-null) and FPR (False Positive Rates, namely the proportion of null coefficients estimated as non-null) for $q^{\star} = 1$ and $q^{\star} = 2$ correspondingly. For each simulation, we considered 9 thresholds ranging from $0.1$ to $0.9$ in the stability selection step. For each threshold, we calculated the maximum difference between TPR and FPR. Then, from
the 9 differences, we took the largest one, which is the best result. It means we did not use the same threshold from one simulation to another.
We considered five different approaches: our method with $q=0$, $q=1$ and $q=2$, classical LASSO for Poisson distribution, and our method
where we took $\boldsymbol{\gamma^{\star}}$ instead of estimating it. More precisely, classical LASSO for Poisson distribution consists in applying the \texttt{glmnet} R package dedicated
to Poisson distribution to the $Y_{i,j,t}$'s for each $t$. We did not compare our method with \texttt{glarma} package because it does not support the multivariate setting.

In Figures \ref{fig1} and \ref{fig3} the closer the maximum difference between TPR and FPR is to 1, the better is the performance of the method.
Our approach with $q=1$ and $q=2$ outperforms classical LASSO and the estimation with $q=0$. We notice that when $J$ is larger, the estimation is better both for $T=50$ and $T=200$. Additionally, the performance for the simulation frameworks with $T=50$ is better than for the ones with $T=200$. In general, our estimation is close to the one with the true value of $\boldsymbol{\gamma^{\star}}$.

\begin{figure}[!htb]
\centering
\includegraphics[scale=0.25]{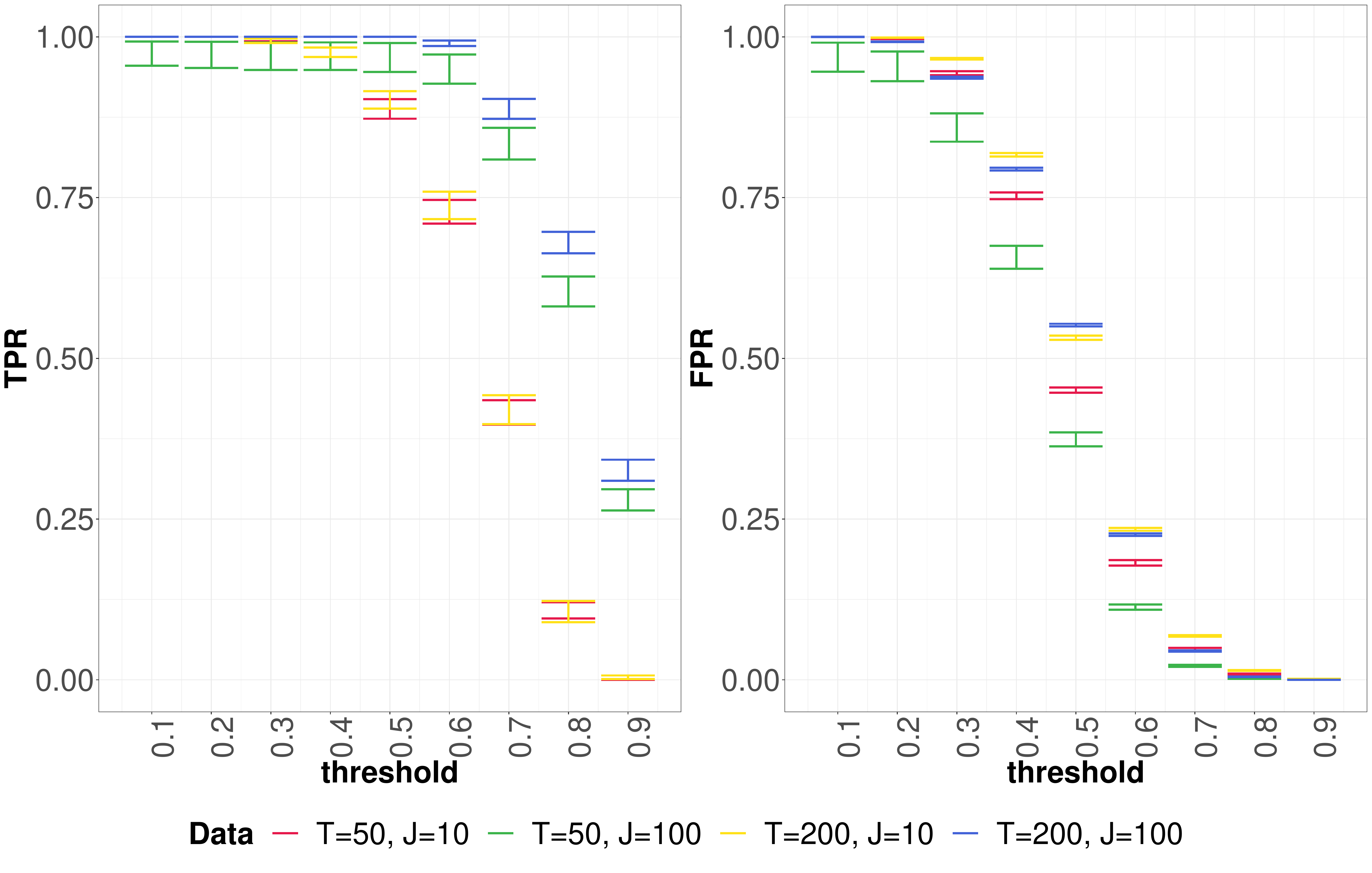}
\caption{Error bars of the TPR and FPR for different thresholds associated to the support recovery of $\boldsymbol{\eta}^\star$ estimated with $q=1$  for 4 different simulation frameworks with respect to the thresholds when $I=3$, $q^{\star}$ = 1, 10 non-null coefficients in $\pmb{\eta^{\star}}$, and 50 simulations.}
\label{fig2}
\end{figure}

\begin{figure}[!htb]
\centering
\includegraphics[scale=0.25]{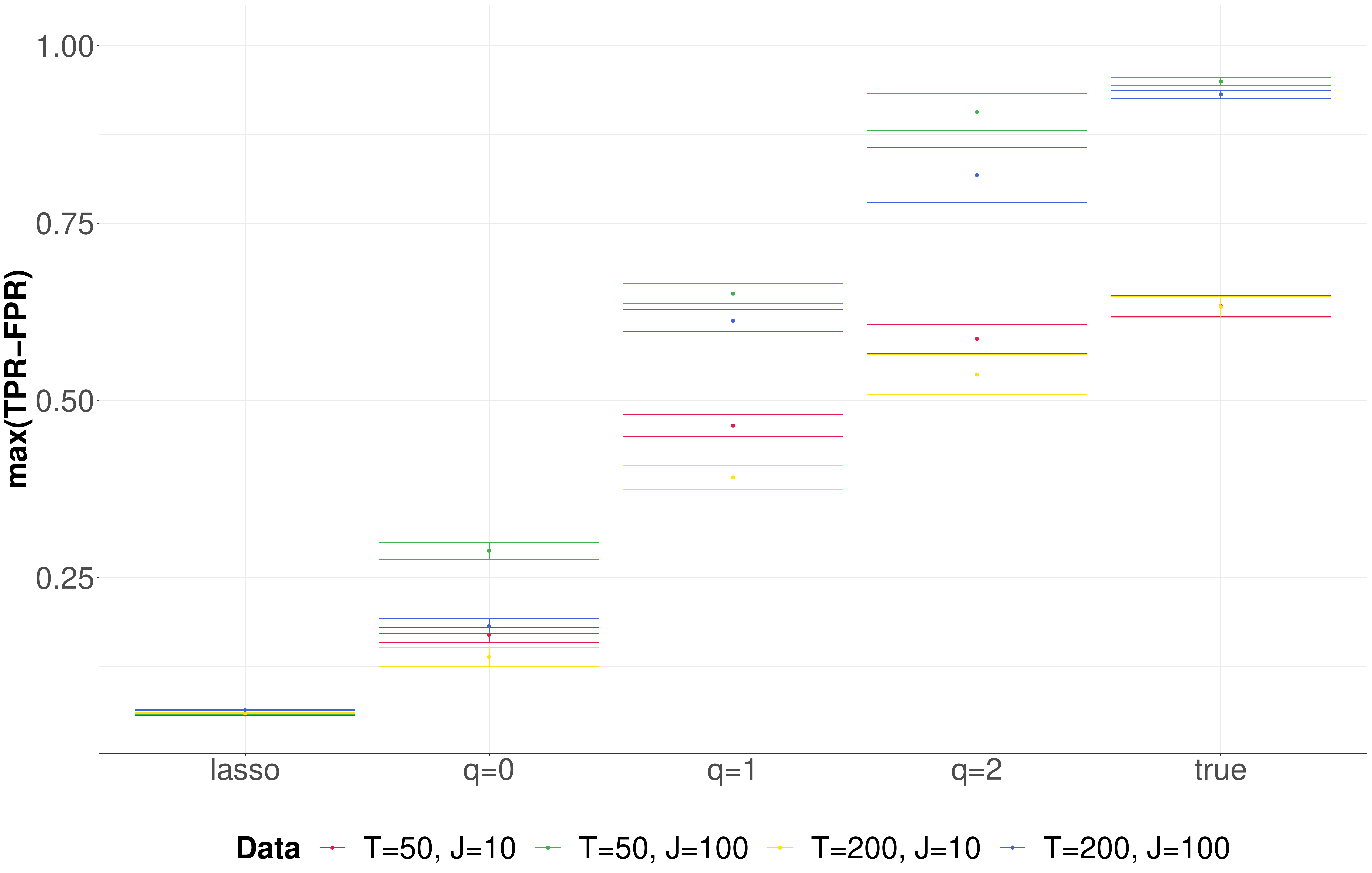}
\caption{Error bars of the maximum difference between TPR and FPR for different thresholds associated to the support recovery of $\boldsymbol{\eta}^\star$  with 5 approaches for 4 simulation frameworks when $I=3$, $q^{\star}$ = 2, 10 non-null coefficients in $\pmb{\eta^{\star}}$, and 50 simulations.}
\label{fig3}
\end{figure}

\begin{figure}[!htb]
\centering
\includegraphics[scale=0.25]{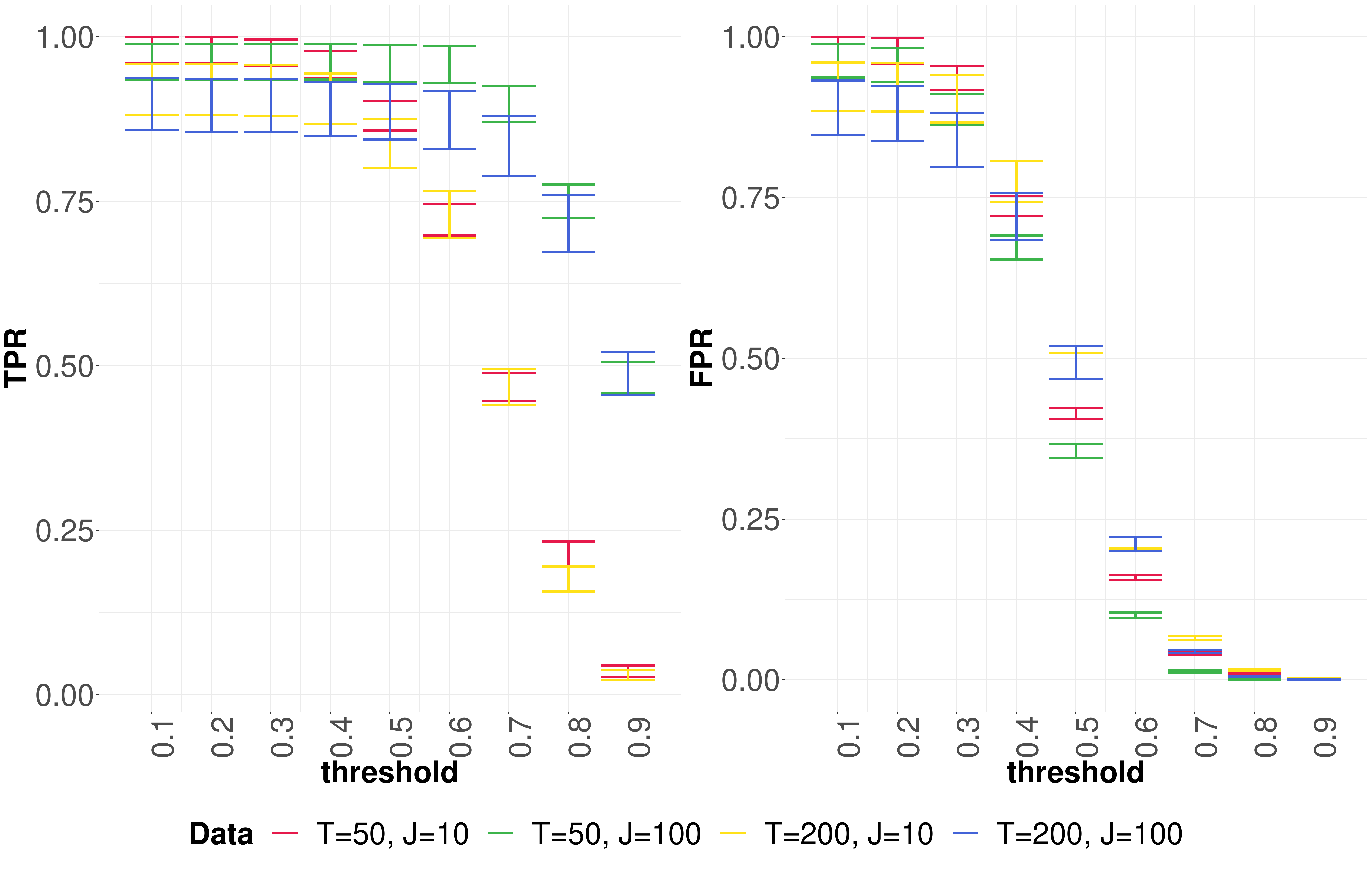}
\caption{Error bars of the TPR and FPR for different thresholds associated to the support recovery of $\boldsymbol{\eta}^\star$ estimated with $q=2$  for 4 different simulation frameworks with respect to the thresholds when $I=3$, $q^{\star}$ = 2, 10 non-null coefficients in $\pmb{\eta^{\star}}$, and 50 simulations.}
\label{fig4}
\end{figure}

Figures \ref{fig2} and \ref{fig4} display the error bars of TPR and FPR of our method with respect to the threshold for $q = 1$ and $q = 2$, respectively. More precisely, the threshold 0.6 achieves a satisfactory trade-off between the TPR and the FPR.
The best trade-offs are achieved for $T=50$ and $J =100$, for both $q = 1$ and $q = 2$.

\clearpage

\paragraph{Sign consistency of the estimation of $\pmb{\eta^{\star}}$}
\label{sec3.2}

\begin{figure}[!htb]
\centering
\includegraphics[scale=0.25]{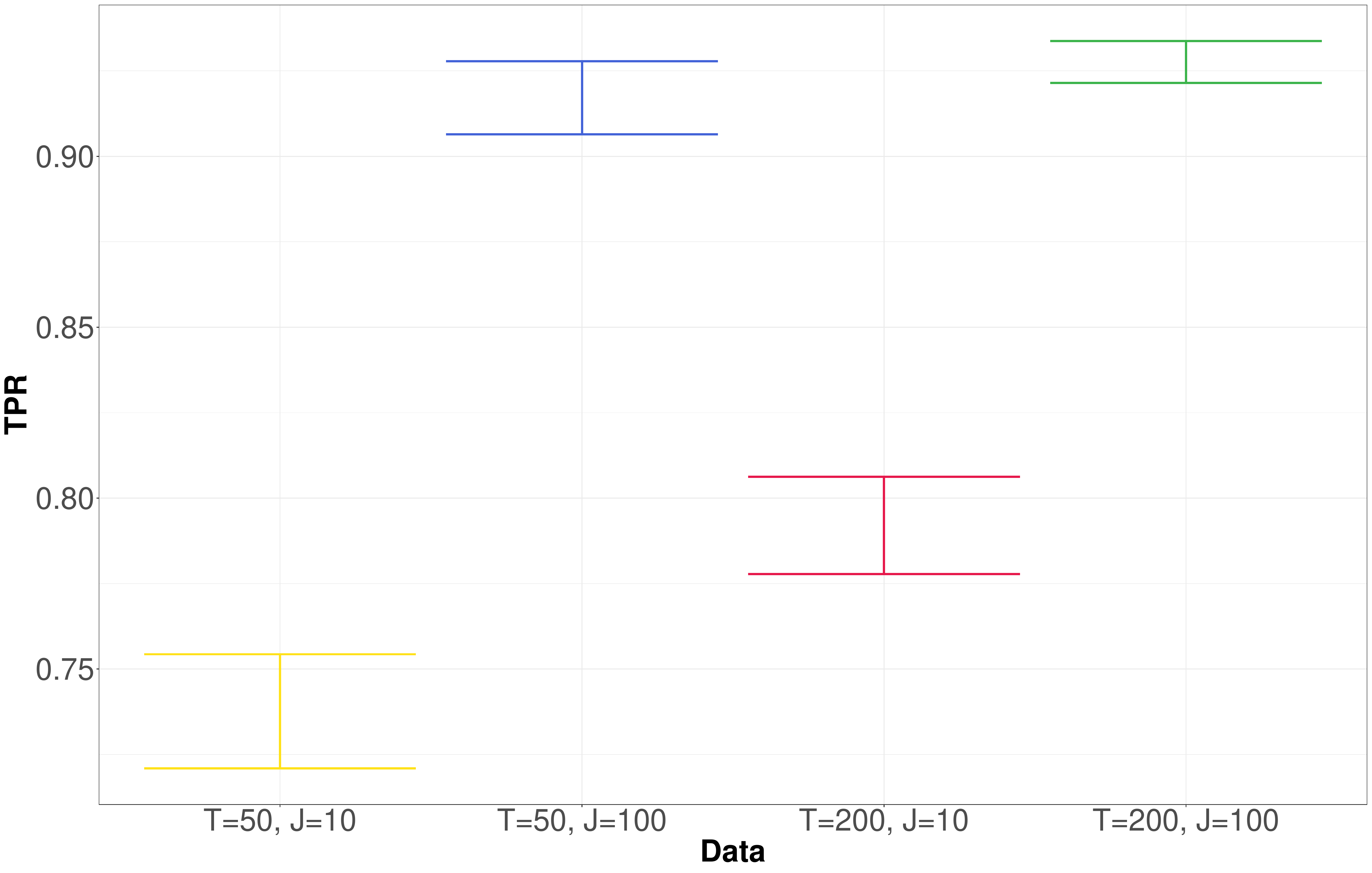}
\caption{Error bars of the TPR of sign recovery of $\boldsymbol{\eta}^\star$  estimated  with $q=1$ for 4 simulation frameworks when $I=3$, $q^{\star}$ = 1, 10 non-null coefficients in $\pmb{\eta^{\star}}$, and 50 simulations.}
\label{fig5}
\end{figure}

\begin{figure}[!htb]
\centering
\includegraphics[scale=0.25]{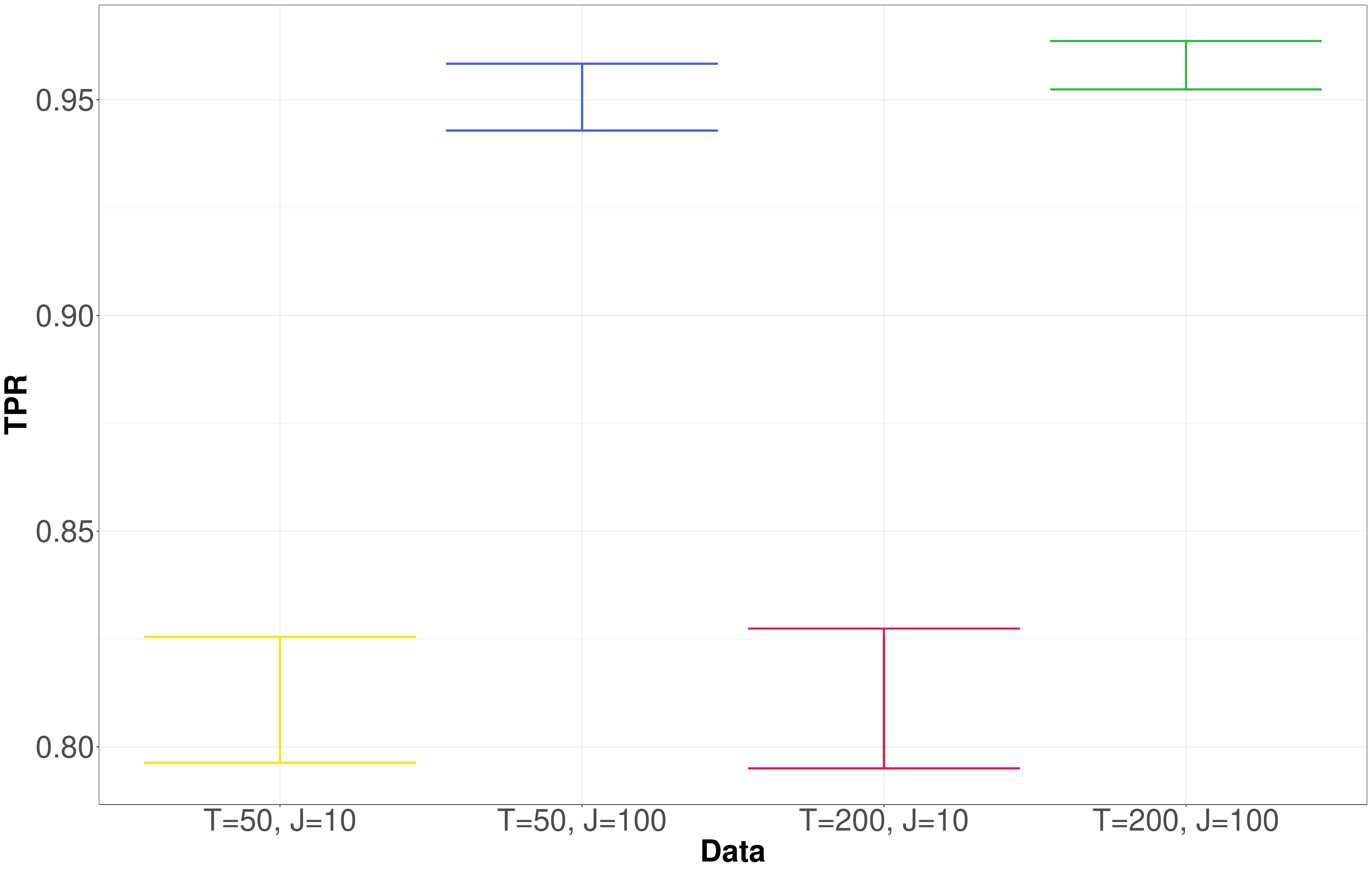}
\caption{Error bars of the TPR of sign recovery of $\boldsymbol{\eta}^\star$  estimated  with $q=2$ for 4 simulation frameworks when $I=3$, $q^{\star}$ = 2, 10 non-null coefficients in $\pmb{\eta^{\star}}$, and 50 simulations.}
\label{fig6}
\end{figure}

In Figures \ref{fig5} and \ref{fig6} we illustrate the TPR of sign recovery of $\boldsymbol{\eta}$. For these figures, we looked at the estimation with the threshold of 0.6. The sign recovery is considered as true positive if for negative (positive) it is estimated with a negative (positive) sign and if 0 is estimated as 0. Here again, we can conclude that the best results are obtained for $J = 100$, similar to the support recovery of $\boldsymbol{\eta^{\star}}$. 


\clearpage

\subsubsection{Estimation of $\pmb{\gamma^{\star}}$}
\label{sec3.3}

\begin{figure}[!htb]
\centering
\includegraphics[scale=0.25]{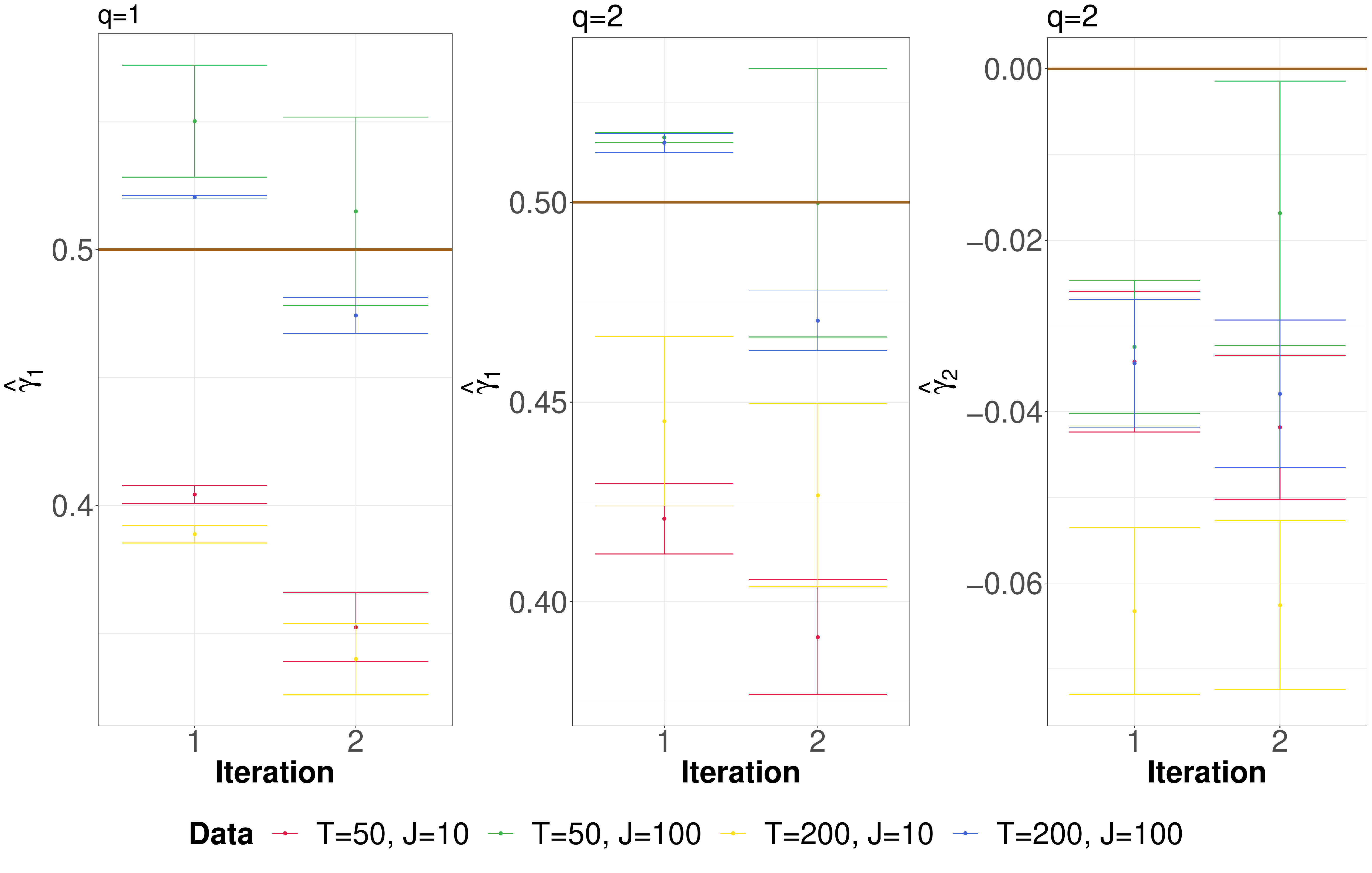}
\caption{Boxplots for the estimations of $\boldsymbol{\gamma}^\star$ in Model \eqref{eq3} for 4 different simulation frameworks when $I=3$, $q^{\star}$ = 1, $\gamma^{\star} = 0.5$, 10 non-null coefficients in $\pmb{\eta^{\star}}$, and 50 simulations obtained by $q=1$ and $q=2$. The horizontal lines correspond to the values of the $\gamma^{\star}_i$’s.}
\label{fig7}
\end{figure}

\begin{figure}[!htb]
\centering
\includegraphics[scale=0.25]{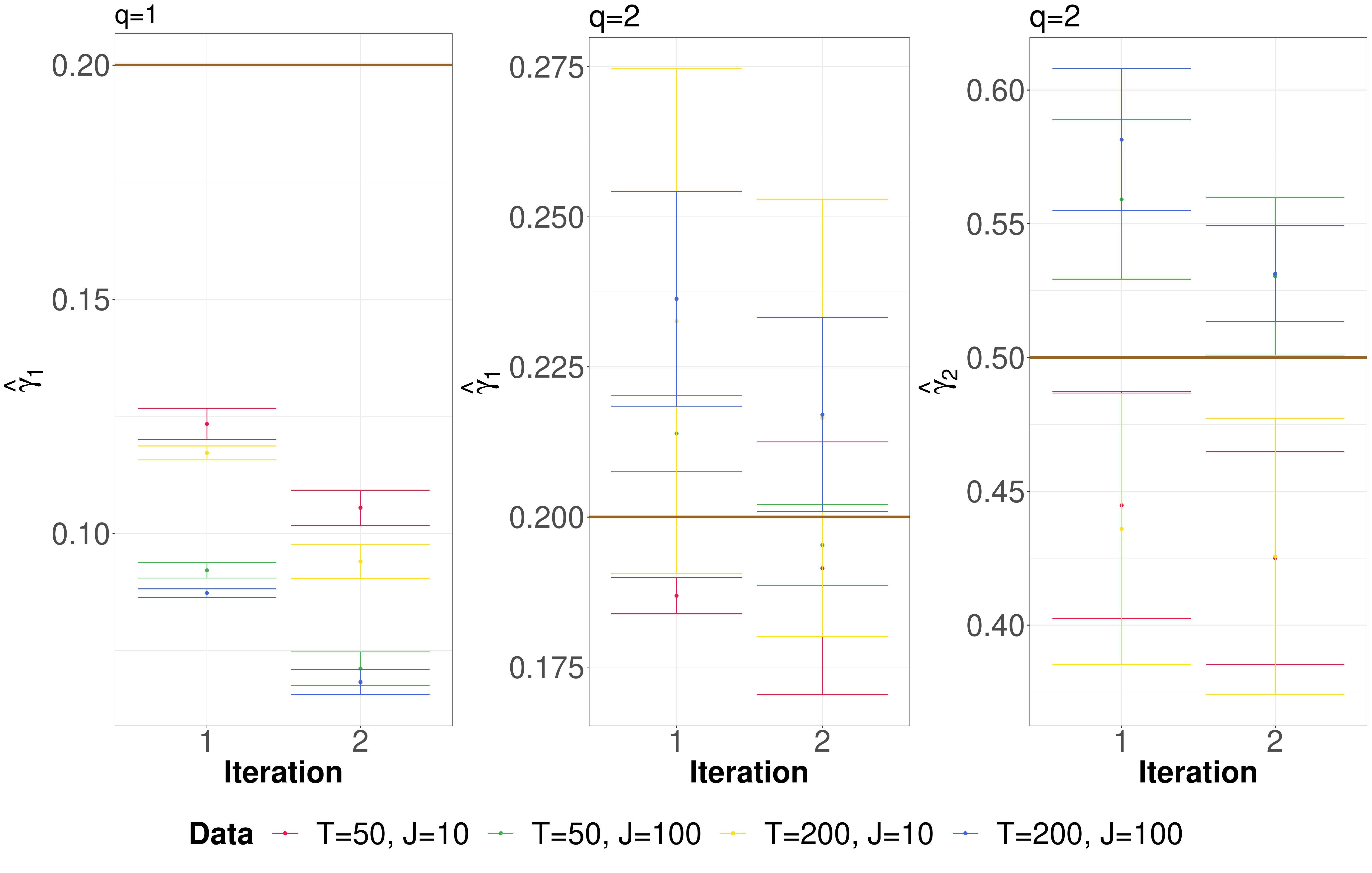}
\caption{Boxplots for the estimations of $\boldsymbol{\gamma}^\star$ in Model \eqref{eq3} for 4 different simulation frameworks when $I=3$, $q^{\star}$ = 2, $\gamma^{\star}_1 = 0.2$, $\gamma^{\star}_1 = 0.5$, 10 non-null coefficients in $\pmb{\eta^{\star}}$, and 50 simulations obtained by $q=1$ and $q=2$. The horizontal lines correspond to the values of the $\gamma^{\star}_i$’s.}
\label{fig8}
\end{figure}

In this section we investigate the performance of the method for the estimation of $\boldsymbol{\gamma^{\star}}$ for the simulation frameworks
of Table \ref{tab1}. In Figures \ref{fig7} (resp. \ref{fig8}), boxplots for the estimations of $\boldsymbol{\gamma}^\star$
in \eqref{eq3} are displayed for $q^{\star} = 1$ (resp. $q^{\star} = 2$). We can see from these figures that when $J=10$, both for $T=50$ and $T=200$, iterating our approach does not improve the results. However, this is not the case for $J =100$: the estimation of $\boldsymbol{\gamma^{\star}}$ improves at the second iteration. In the Appendix \ref{app:num}, we present additional figures for the settings $T=50$ with $J =10$ and $J =100$, and 10 iterations. These plots justify that for a small value of $J$ iterating the method does not improve the estimation, whereas for a large value of $J$  the estimation stabilises and results become better.

\clearpage

\subsection{Numerical performance}
\begin{figure}[!htb]
\centering
\includegraphics[scale=0.25]{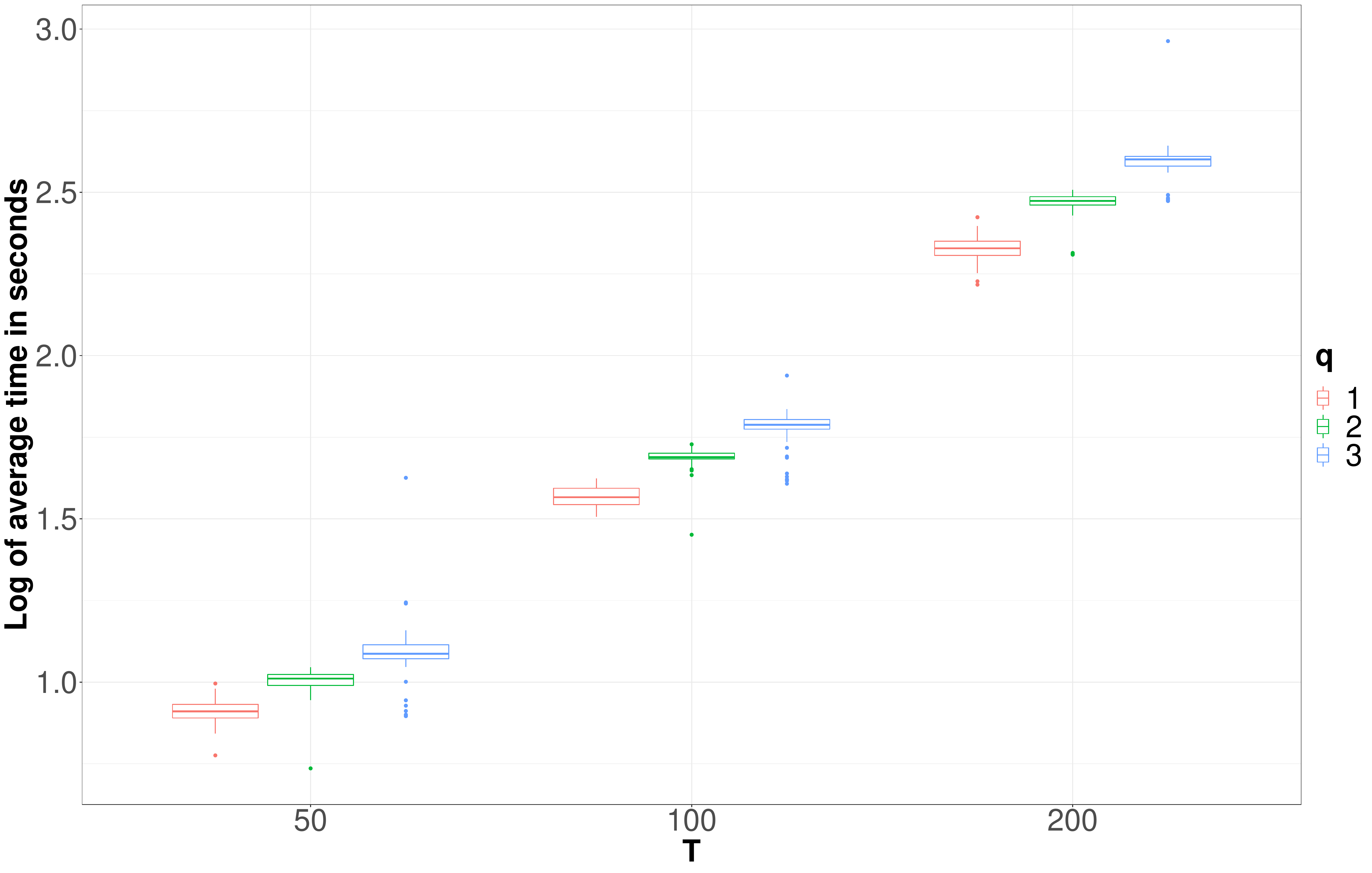}
\caption{Boxplots of the $\log_{10}$ computational times in seconds  in the case where $I=3$, $J = 10 $, $q = q^\star$, 10 non-null coefficients in $\pmb{\eta^{\star}}$, and different values of $T$ and $q^\star$, a given threshold and one iteration. We performed 50 simulations.}
\label{fig:time}
\end{figure}

Figure \ref{fig:time} displays the means of the computational times of our approach implemented in the R package \texttt{multiGlarmaVarSel}
for different simulation frameworks.
The timings were obtained on a workstation with 32GB of RAM and Intel Core i7-9700 (3.00GHz) CPU.
We can see from this figure that the computational time goes from 10 seconds to 5 minutes to process the data for a given threshold and one iteration, when we increase $T$ from $50$ to $200$ and when $q=1$, 2 or 3. 

\clearpage

\section{Application to RNA-Seq data}
\label{sec4}
\subsection{Biological context and modelling}

In order to address the issue related to the influence of the thermal environment of the mother plant on the germination potential of the progeny at the scale of translational regulation, the model plant, Arabidopsis thaliana (Col-0 accession), was cultivated  under three temperature regimes (Low, $14–16^\circ$C; Medium, $18–22^\circ$C ; Elevated, $25–28^\circ$C) under a long-day photoperiod. As described in the introduction, the lower the cultivation temperature, the deeper the dormancy of the harvested seeds~\cite{penfield:macgregor:2016}. Seeds produced under the three temperature regimes were placed under germination conditions at low temperature ($10^\circ $C)  to avoid thermo-dormancy. After 72 hours of imbibition, before the first radicle protrusion, the seeds were collected for molecular analyses. Polysome profiling of Arabidopsis seeds was performed as described by~\cite{basbouss:etal:2015}. The purified polysomal mRNAs and non-polysomal mRNAs were analysed by RNA-sequencing~\cite{stark:etal:2019}.

The data consists of 26725 gene expressions observed in 3 conditions of temperature with 3 replicates for 5 chromosomes and two mRNA populations
(polysomal and non-polysomal).
Since the gene expresssions are integer values and since there
may be some dependence between them we modeled this data by using Model (\ref{eq1})--(\ref{eq4}). In this model,
$Y_{i,j,t}$ is a random variable describing the expression of the $j$th replication of gene $t$ in temperature $i$ with $I=3$, $n_i=J=3$ for all $i$ and $T=26725$.
Moreover, $\eta_{i,t}^\star$ corresponds to the effect of temperature $i$ on gene $t$.

In this framework, where $I$ and $J$ are very small, according to the numerical results obtained in Section \ref{sec3}, the value of $T$ has to be reduced to obtain
satisfactory statistical performance. This is the reason why we preprocessed the data as follows.
For each mRNA population and each of the five chromosomes, we used a one-way ANOVA GLM with Poisson distribution to identify the genes on which the conditions have an influence.
We kept the genes for which the
$p$-value of the corresponding $t$-test is smaller than $1/{T_{c,pop}}$ where $T_{c,pop}$ is the number of genes present in the chromosome $c$ and in the
mRNA population $pop$ ($T_{c,pop}$ ranges from 4074 to 7003). With this filtering, the new values of $T_{c,pop}$ for each mRNA population are given in Table \ref{tab2} where ``Non-poly'' (resp. ``Poly'') refers to non-polysomal (resp. polysomal).

\subsection{Results obtained with the \texttt{multiGlarmaVarSel} R package}

This section provides the results obtained by applying our methodology to each of the five chromosomes of each mRNA population.
Since $I$ and $J$ are very small in this application, we only focus on $q=1$. Based on the results obtained in \ref{sec3.3}, we only
ran one iteration of our procedure.

The estimation of $\gamma_1^\star$ for the polysomal and non-polysomal mRNA populations and the different chromosomes
are given in Table \ref{tab4}. We can see from this table that the estimations are similar for the two populations
except for chromosomes 3 and 4.


\begin{table}[H]
\centering
\begin{tabular}{|c|c|c|}
\hline
Chromosome & $\widehat{\gamma}_1$ for non-poly population & $\widehat{\gamma}_1$ for poly population  \\ \hline
  1	& -0.00369	        & 0.06234          \\ \hline
2	& 0.05617		& 0.04312	   \\ \hline
3	& 0.03314		& 0.79662           \\ \hline
4	& 0.21809		& 0.09473          \\ \hline
5	& 0.00574		& 0.00159         \\ \hline
\end{tabular}
\caption{Estimation of $\gamma_1^{\star}$ for the 5 chromosomes and the two mRNA populations.}
\label{tab4}
\end{table}

Table \ref{tab2} provides the number of genes selected by our procedure in the two mRNA populations for the
different chromosomes. We can see from this table that the number of selected genes that are common in the two mRNA populations
ranges from 1 to 9 and is the highest for the first chromosome.

\begin{table}[H]
\centering
\begin{tabular}{|c|c|c|c|c|c|}
\hline
  Chromosome & $T_{c,\textrm{Non-poly}}$ & Selected genes in  & $T_{c,\textrm{Poly}}$ & Selected genes in  & Intersection   \\ 
             &                         & non-poly population            &          &  poly population        &               \\ \hline
1	& 70	& 59	& 25          & 20 &  9        \\ \hline
2	& 59	& 44 	& 16          & 12   &    6    \\ \hline
3	& 37	& 31   & 4          & 1    &     1   \\ \hline
4	& 41	& 25   & 18          & 12  &  4  \\ \hline
5	& 43	& 39   & 9          & 7    & 1    \\ \hline
\end{tabular}
\caption{Number of genes selected by our procedure with $q=1$ in the two mRNA populations as well as those that are common in the both (Intersection column).}
\label{tab2}
\end{table}

\begin{figure}[!htb]
\centering
\includegraphics[scale=0.25]{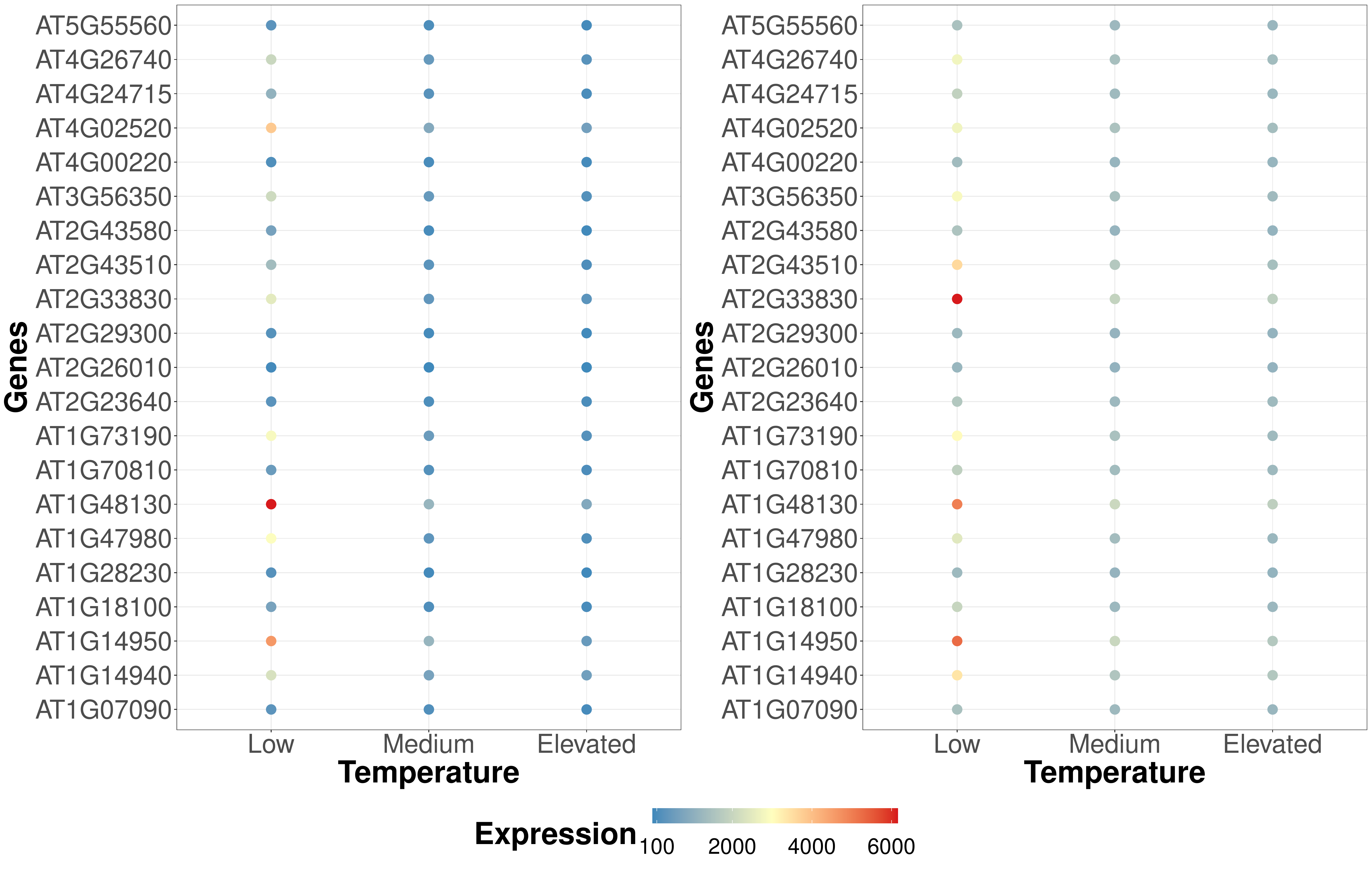}
\caption{Average gene expression values of 3 replications at each temperature condition
  (Low, Medium, Elevated) for the non-polysomal (resp. polysomal) population on the left (resp. right). The genes displayed in this figure are obtained by our selection procedure and are common to both populations.}
\label{fig:application}
\end{figure}

Figure \ref{fig:application} displays the average gene expression values of the 3 replications for each temperature condition
(Low, Medium, Elevated). The genes displayed in this figure are obtained by our selection procedure and are common to the two populations
(polysomal and non-polysomal).
We can see from this figure that the temperature conditions may have a different impact on the expression of the genes.
This is the case, for instance, for AT1G48130, AT2G33830 and AT1G14950, on which the low temperature has a positive effect on their expression. Figures \ref{fig:non_poly_1}, \ref{fig:non_poly_2}, \ref{fig:non_poly_3_4}, \ref{fig:non_poly_5} and \ref{fig:poly} in Appendix \ref{app:app} display the average gene expression values of all the selected genes for non-polysomal  and polysomal populations.

\FloatBarrier

\subsection{Biological results}

\begin{figure}[!htb]
\centering
\includegraphics[scale=0.35]{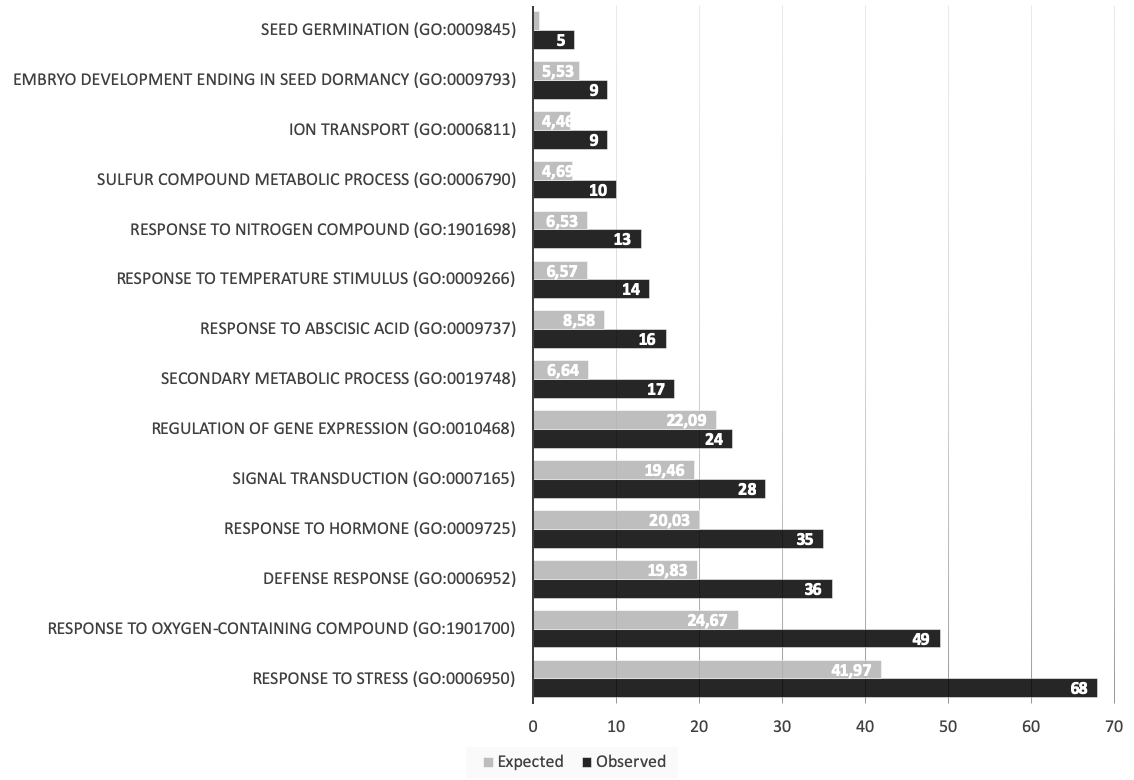}
\caption{Gene ontology (GO) term enrichment analysis of the selected 229 genes based on polysomal-associated mRNA differentially accumulated in germinating seeds produced beforehand under different temperature regimes. Data from PANTHER overrepresentation test (http://www.geneontology.org); Arabidopsis thaliana GO database released March $22^{\text{nd}}$ 2022 (DOI: 10.5281/zenodo.6399963). Black bars: observed gene number in the selection; grey bars: expected gene number from the reference Arabidopsis genome.}
\label{fig:gene_ontology}
\end{figure}

The results allowed the selection of 198 non-polysomal mRNAs and 52 polysomal mRNAs accumulated differentially in the germinating seeds under the influence of the growth temperature of the mother plant. Overall it concerns 229 genes, and 21 are shared in both populations of mRNA. 
It is worth noting that several genes previously described to be involved in the control of seed dormancy and germination were selected by the present statistical method, such as RDO5 (AT4G11040,~\cite{xiang:etal:2014}), DRM2 (AT2G33830,~\cite{iwasaki:etal:2019}), DOG1-like 3 (AT4G18690,~\cite{bentsink:etal:2006}), MFT (AT1G18100,~\cite{xi:etal:2010}), XERICO (AT2G04240,~\cite{ko:etal:2006}) or HAI3 (AT2G29380,~\cite{nishimura:etal:2018}). Thus, these genes also seem to be involved in the modulation of germination potential induced by the thermal environment of the mother plant. A gene ontology (GO) analysis from the 229 genes revealed that the top 5 of the biological processes affected were the response to stress (GO:0006950), response to oxygen-containing compound (GO:1901700), defence response (GO:0006952), response to hormone (GO:0009725) and signal transduction (GO:0007165) (Figure \ref{fig:gene_ontology}). This is a pioneering observation showing that the environment of the mother plant not only influences the germinative potential of offspring seeds through hormonal and redox regulation~\cite{shu:etal:2016, bailly:2019}, but also the ability of germinating seeds to cope with biotic and abiotic stresses. Interestingly, 23 transcription factors were selected by the statistical approach (Table \ref{biological_interpretation}). These genes could represent key regulators of the modulation of seed physiological quality in response to various types of biotic and environmental stress during seed production and/or during germination. These results open the door for further research addressing the question of the control of mRNA metabolism during seed germination, notably concerning the selectivity of translational control. The germinating seed is undoubtedly a relevant biological model for exploring the precise mechanisms of combined transcriptional and translational regulation related to gene expression ending with the production of functional protein.

\FloatBarrier

\newpage

{\fontsize{7}{9}\selectfont
\renewcommand*{\arraystretch}{1.4}
\begin{longtable}{P{1.8cm}P{1.8cm}p{4.5cm}p{2.3cm}P{2cm}}
\hline
\textbf{TF Family Name} & \textbf{AGI} & \textbf{Gene TAIR Curator Summary} & \textbf{Gene TAIR Short Description} & \textbf{Gene TAIR Aliases} \\
\hline
AP2-EREBP & AT1G28360 & Encodes a member of the ERF (ethylene response factor) subfamily B-1 of ERF/AP2 transcription factor family (ERF12). The protein contains one AP2 domain. & ERF domain protein 12  & ATERF12, ERF12 \\
AP2-EREBP & AT1G75490 & Encodes a member of the DREB subfamily A-2 of ERF/AP2 transcription factor family. The protein contains one AP2 domain.  & Integrase-type DNA-binding superfamily protein  & DREB2D, ERF049 \\
AP2-EREBP & AT3G16770    & Encodes a member of the ERF (ethylene response factor) subfamily B-2 of the plant specific ERF/AP2 transcription factor family (RAP2.3).  The protein contains one AP2 domain. Overexpression of this gene in tobacco BY-2 cells confers resistance to H2O2 and heat stresses. Overexpression in Arabidopsis causes upregulation of PDF1.2 and GST6. It is part of the ethylene signalling pathway and is predicted to act downstream of EIN2 and CTR1, but not under EIN3. & Ethylene-responsive element binding protein                                            & ATEBP, EBP, ERF72, RAP2.3  \\
AP2-EREBP & AT5G53290 & Encodes a member of the ERF (ethylene response factor) subfamily B-5 of ERF/AP2 transcription factor family. The protein contains one AP2 domain. CRF proteins relocalize to the nucleus in response to cytokinin. & Cytokinin response factor 3 & CRF3  \\
AP2-EREBP  & AT5G65510 & Encodes one of three PLETHORA transcription factors required to maintain high levels of PIN1 expression at the periphery of the meristem and modulate local auxin production in the central region of the SAM which underlies phyllotactic transitions. & AINTEGUMENTA-like 7 & AIL7, PLT7 \\
C2C2-Gata  & AT3G51080  & Encodes a member of the GATA factor family of zinc finger transcription factors. & GATA transcription factor 6 & GATA6 \\
C2H2  & AT2G41940  & Encodes a zinc finger protein containing only a single zinc finger. Involved in GA and cytokinin signal integration. & Zinc finger protein 8  & ZFP8 \\
C2H2 & AT3G07940 & Calcium-dependent ARF-type GTPase activating protein family & Calcium-dependent ARF-type GTPase activating protein family & AtGAP \\
C2H2 & AT5G07500  & Encodes an embryo-specific zinc finger transcription factor required for heart-stage embryo formation. & Zinc finger C-x8-C-x5-C-x3-H type family protein  & AtTZF6, PEI1, TZF6  \\
C2H2  & AT5G43540    & Encodes a protein containing a zinc finger, C2H2-type domain. & C2H2 and C2HC zinc fingers superfamily protein & ZF-C2H2-type \\
C3H  & AT2G04240   & Encodes a small protein with an N-terminal trans-membrane domain and a RING-H2 zinc finger motif located at the C-terminus.  Gene expression is induced by salt and osmotic stress. Transcript levels are induced by DELLA proteins and repressed by gibberellic acid. Involved in ABA metabolism. & RING/U-box superfamily protein  & XERICO   \\
CCAAT-HAP2  & AT3G14020    & Encodes a nuclear factor Y A (NF-YA), a highly conserved transcription factor presented in all eukaryotic organisms  & Nuclear factor Y, subunit A6 & NF-YA6  \\
CCAAT-HAP5  & AT5G27910  & Encodes a nuclear factor Y C (NF-YC), a highly conserved transcription factor presented in all eukaryotic organisms & Nuclear factor Y, subunit C8 & NF-YC8 \\
G2-like   & AT2G20570  & Encodes GLK1, Golden2-like 1, one of a pair of partially redundant nuclear transcription factors that regulate chloroplast development in a cell-autonomous manner. GLK2, Golden2-like 2, is encoded by At5g44190. GLK1 and GLK2 regulate the expression of the photosynthetic apparatus. & GBF's pro-rich & ATGLK1, GLK1, GPRI1  \\
G2-like                 & AT4G04580    & Encodes a protein containing Myb, DNA-binding domain & Homeodomain-like superfamily protein & MYB-TF  \\
G2-like  & AT5G16560  & Encodes a KANADI protein (KAN) that regulates organ polarity in Arabidopsis. KAN encodes a nuclear-localised protein in the GARP family of putative transcription factors. Together with KAN2, this gene appears to be involved in the development of the carpel and the outer integument of the ovule. Along with KAN2 and KAN4, KAN1 appears to be required for proper regulation of PIN1 in early embryogenesis. & Homeodomain-like superfamily protein & KAN, KAN1  \\
Homeobox  & AT1G52150    & Member of the class III HD-ZIP protein family. Contains homeodomain and leucine zipper domain. Critical for vascular development and negatively regulates vascular cell differentiation. & Homeobox-leucine zipper family protein / lipid-binding START domain-containing protein & ATHB-15, ATHB15, CNA, ICU4 \\
MADS  & AT5G51870  & Encodes a MADS-box transcription factor involved in floral transition. & AGAMOUS-like 71  & AGL71  \\
MADS  & AT5G65070  & Encodes MADS-box containing FLC paralog. Five splice variants have been identified but not characterised with respect to expression patterns and/or differing function. Overexpression of the gene in the Landsberg ecotype leads to a delay in flowering, transcript levels of MAF4 are reduced after a 6 week vernalization. & K-box region and MADS-box transcription factor family protein  & AGL69, FCL4, MAF4\\
NAC   & AT1G52890   & Encodes a NAC transcription factor whose expression is induced by drought, high salt, and abscisic acid. This gene binds to ERD1 promoter in vitro. & NAC domain containing protein 19  & ANAC019, ANAC19, NAC019  \\
NAC  & AT1G69490    & Encodes a member of the NAC transcription factor gene family.  It is expressed in floral primordia and upregulated by AP3 and PI.  Its expression is associated with leaf senescence. The mRNA is cell-to-cell mobile. & NAC-like, activated by AP3/PI & ANAC029, ATNAP, NAP \\
NAC  & AT2G27300  & NTL8 is a membrane-associated NAC transcription factor that binds both TRY and TCL1. Overexpression results in fewer trichomes. & NTM1-like 8 & ANAC040, NTL8  \\
NLP & AT3G59580 & NIN-LIKE PROTEIN 9 & Plant regulator RWP-RK family protein & NLP9  \\  
\hline & \\[-1.5ex]
\caption{List of 23 transcription factors selected by statistical analysis based on polysomal-associated mRNA differentially accumulated in germinating seeds produced beforehand under different temperature regimes. }
\label{biological_interpretation}
\end{longtable}}

\appendix
\section{Appendix}
\subsection{Computation of the first and second derivatives of $W_t$ defined in \eqref{eq:Wijt}}
\label{app}
\subsubsection{Computation of the first derivatives of $W_t$ }
\label{app1}

By the definition of $W_t$ given in  (\ref{eq:Wijt}), we have
\begin{equation*}
\frac{\partial W_{i,j,t}}{\partial \pmb{\delta}} = \frac{\partial \eta_{i,t}}{\partial \pmb{\delta}} + \frac{\partial Z_{i,j,t}}{\partial \pmb{\delta}}.
\end{equation*}
For all $i_0 \in \{ 1, \dots, I\}$ and $t_0 \in \{ 1, \dots, T\}$ we have
\begin{equation*}
\begin{split}
\frac{\partial W_{i,j,t}}{\partial \eta_{i_0, t_0}}  &= \frac{\partial}{\partial \eta_{i_0, t_0}} \Big( \eta_{i,t} + Z_{i,j,t} \Big) = \frac{\partial \eta_{i,t}}{\partial \eta_{i_0, t_0}} + \sum_{k=1}^{q \wedge (t-1)}  \gamma_k \frac{\partial E_{i,j,t-k} }{\partial  \eta_{i_0, t_0}} \\
&= \frac{\partial \eta_{i,t}}{\partial \eta_{i_0, t_0}}  - \sum_{k=1}^{q \wedge (t-1)} \gamma_k Y_{i,j,t-k} \frac{\partial W_{i,j,t-k}}{\partial \eta_{i_0, t_0}} \exp(-\eta_{i, t-k} - Z_{i,j,t-k}) \\
& = \frac{\partial \eta_{i,t}}{\partial \eta_{i_0, t_0}}  - \sum_{j=k}^{q \wedge (t-1)}  \gamma_k (1 + E_{i,j,t-k}) \frac{\partial W_{i,j,t-k}}{\partial \eta_{i_0, t_0}},
\end{split}
\end{equation*}
where $E_{i,j,t} = 0$ for any $t \leq 0$.

For all $q_0 \in \{ 1, \dots, q\}$
\begin{equation*}
\begin{split}
\frac{\partial W_{i,j,t}}{\partial \gamma_{q_0}}  &= \frac{\partial}{\partial \gamma_{q_0}} \Big( \eta_{i,t} + Z_{i,j,t} \Big) = \frac{\partial \eta_{i,t}}{\partial \gamma_{q_0}} + \frac{\partial}{\partial \gamma_{q_0}}  \sum_{k=1}^q \gamma_k E_{j,t-k}^{(i)} = E_{j, t-q_0}^{(i)} + \sum_{k=1}^{q \wedge (t-1)} \gamma_k  \frac{\partial E_{i,j,t-k}}{\partial \gamma_{q_0}} \\
& = E_{i,j,t-q_0} - \sum_{k=1}^{q \wedge (t-1)} \gamma_k Y_{i,j,t-k}  \frac{\partial W_{i,j,t-k}} {\partial \gamma_{q_0}}  \exp(-W_{i,j,t-k}) \\
& = E_{i,j,t-q_0} - \sum_{k=1}^{q \wedge (t-1)} \gamma_k (1 + E_{i,j,t-k}) \frac{\partial W_{i,j,t-k}}{\partial \gamma_{q_0}},
\label{app_eq_3}
\end{split}
\end{equation*}
where we used the fact that $E_{i,j,t-q_0} = 0$ for any $t \leq 0$ . 

We obtain the first derivatives of $W_{i,j,t}$ from the following recursive expressions. For all $i_0 \in \{ 1, \dots, I\}$ and $t_0 \in \{ 1, \dots, T\}$
\begin{equation*}
\begin{split}
& \frac{\partial W_{i,j,1}}{\partial \eta_{i_0, t_0}} =  \frac{\partial \eta_{i,1}}{\partial \eta_{i_0, t_0}} =     
\begin{cases}
      1, \quad \text{if } i = i_0 \quad \text{and } t_0 = 1 \\
      0, \quad \text{otherwise}
    \end{cases}\, , \\
& \frac{\partial W_{i,j,2}}{\partial \eta_{i_0, t_0}} =  \frac{\partial \eta_{i,2}}{\partial \eta_{i_0, t_0}} -  \gamma_1 (1 + E_{i,j,1})   \frac{\partial W_{i,j,1}}{\partial \eta_{i_0, t_0}} \\
&= \begin{cases}
      1, \quad \text{if } i = i_0 \quad \text{and } t_0 = 2 \\
      -\gamma_1 (1 + E_{i,j,1}), \quad \text{if } i = i_0 \quad \text{and } t_1 = 1 \\ 
      0, \quad \text{otherwise}
    \end{cases}\, , \\
\label{app_eq_4}
\end{split}
\end{equation*}

In the same way, for all $q_0 \in \{1, \dots, q\}$ we have 
\begin{equation*}
\begin{split}
& \frac{\partial W_{i,j,1}}{\partial \gamma_{q_0}} = 0, \\
& \frac{\partial W_{i,j,2}}{\partial \gamma_{q_0}} = E_{i,j,2-q_0}, \\
& \frac{\partial W_{i,j,3}}{\partial \gamma_{q_0}}= E_{i,j,3-q_0} - \gamma_1 (1 + E_{i,j,2}) \frac{\partial W_{i,j,2}}{\partial \gamma_{q_0}},
\label{app_eq_8}
\end{split}
\end{equation*}
and so on. Note that 
\begin{equation*}
\begin{split}
& W_{i,j,1}= \eta_{i, 1} + Z_{i,j,1} = \eta_{i, 1} + \sum_{k=1}^q \gamma_k E_{i,j,1-k} = \eta_{i, 1},\\
& E_{i,j,1} = Y_{i,j,1}  \exp(-W_{i,j,1}) - 1 = Y_{i,j,1} \exp(-\eta_{i, 1}) - 1,\\
& W_{i,j,2} = \eta_{i, 2} + Z_{i,j,2} = \eta_{i, 2} + \sum_{k=1}^q \gamma_k E_{i,j,2-k}= \eta_{i, 2} + \gamma_1 E_{i,j,1}, \\
& E_{i,j,2} = Y_{i,j,2}  \exp(-W_{i,j,2}) - 1 = Y_{i,j,2} \exp(-\eta_{i, 2} - \gamma_1 E_{i,j,1}) - 1.\\
\end{split}
\end{equation*}

\subsubsection{Computation of the second derivatives of $W_t$ }
\label{app2}

For all $i_0, i_1 \in \{ 0, \dots, I \}$ and $t_0, t_1 \in \{ 1, \dots, T \}$ 
\begin{equation*}
\begin{split}
\frac{\partial^2 W_{i,j,t}}{\partial \eta_{i_0, t_0} \partial \eta_{i_1, t_1}}  &= \frac{\partial}{\partial \eta_{i_1, t_1}} \Bigg\{ \frac{\partial \eta_{i, t}}{\partial \eta_{i_0, t_0}} - \sum_{k=1}^{q \wedge (t-1)} \gamma_k(1+E_{i,j,t-k}) \frac{\partial W_{i,j,t-k}}{\partial \eta_{i_0, t_0} } \Bigg\} \\
& = - \sum_{k=1}^{q \wedge (t-1)} \gamma_k \frac{\partial E_{i,j,t-k}}{\partial \eta_{i_1, t_1}} \frac{\partial W_{i,j,t-k}}{\partial \eta_{i_0, t_0} } - \sum_{k=1}^{q \wedge (t-1)} \gamma_k (1+E_{i,j,t-k})  \frac{\partial^2 W_{i,j,t-k}}{\partial \eta_{i_0, t_0} \partial \eta_{i_1, t_1}} \\
& = \sum_{k=1}^{q \wedge (t-1)} \gamma_k (1+E_{i,j,t-k}) \frac{\partial W_{i,j,t-k}}{\partial \eta_{i_0, t_0}} \frac{\partial W_{i,j,t-k}}{\partial \eta_{i_1, t_1}} - \sum_{k=1}^{q \wedge (t-1)}  \gamma_k (1+E_{i,j,t-k}) \frac{\partial^2 W_{i,j,t-k}}{\partial \eta_{i_0, t_0} \partial  \eta_{i_1, t_1}}. \\
\label{app_eq_9}
\end{split}
\end{equation*}

For all $q_0, q_1 \in \{1, \dots, q \}$
\begin{equation*}
\begin{split}
\frac{\partial^2 W_{i,j,t}}{\partial \gamma_{q_0} \partial \gamma_{q_1}}   &= \frac{\partial E_{i,j,t-q_0}}{\partial \gamma_{q_1}}-(1+E_{i,j,t-q_1}) \frac{\partial W_{i,j,t-q_1}}{\partial \gamma_{q_0}} - \sum_{k=1}^{q \wedge (t-1)} \gamma_k\Bigg\{ \frac{\partial W_{i,j,t-k}}{\partial \gamma_{q_0}} \frac{\partial E_{i,j,t-k}}{\partial \gamma_{q_1}} \\ 
&+(1+E_{i,j,t-k}) \frac{\partial^2 W_{i,j,t-k}}{\partial \gamma_{q_0} \partial \gamma_{q_1}}  \Bigg\} = -(1 + E_{i,j,t-q_0}) \frac{\partial W_{j, t-q_0}}{\partial \gamma_{q_1}} - (1 + E_{i,j,t-q_1}) \frac{\partial W_{i,j,t-q_1}}{\partial \gamma_{q_0}}  \\ 
& - \sum_{k=1}^{q \wedge (t-1)} \gamma_k \Bigg\{ -(1+E_{i,j,t-k}) \frac{\partial W_{i,j,t-k}}{\partial \gamma_{q_0}} \frac{\partial W_{i,j,t-k}}{\partial \gamma_{q_1}} + (1+E_{i,j,t-k}) \frac{\partial^2 W_{i,j,t-k}}{\partial\gamma_{q_0} \partial \gamma_{q_1}} \Bigg\}.
\label{app_eq_11}
\end{split}
\end{equation*}

To obtain the second derivatives of $W_t$ we use the following recursive expressions for all $i_0,i_1 \in \{ 0, \dots, I\}$ and $t_0, t_1 \in \{ 1, \dots, T \}$ 
\begin{equation*}
\begin{split}
& \frac{\partial^2 W_{i,j,1}}{\partial \eta_{i_0, t_0} \partial \eta_{i_1, t_1}} = 0, \\
& \frac{\partial^2 W_{i,j,2}}{\partial \eta_{i_0, t_0} \partial \eta_{i_1, t_1}} = 
\begin{cases}
       \gamma_1 (1 + E_{i,j,1}), \quad \text{if } i = i_0= i_1 \quad \text{and } t_0 = t_1 = 1 \\
       0, \quad \text{otherwise}
    \end{cases}\, .\\
\label{app_eq_12}
\end{split}
\end{equation*}
We also have that for all $q_0, q_1 \in \{1, \dots, q \}$
\begin{equation*}
\begin{split}
& \frac{\partial^2 W_{i,j,1}}{\partial \gamma_{q_0} \partial \gamma_{q_1}} = 0, \\
& \frac{\partial^2 W_{i,j,2}}{\partial \gamma_{q_0} \partial \gamma_{q_1}} = 0,\\
\label{app_eq_13}
\end{split}
\end{equation*}
and so on.

\subsection{Additional numerical experiments}\label{app:num}

\begin{figure}[!htb]
\centering
\includegraphics[scale=0.25]{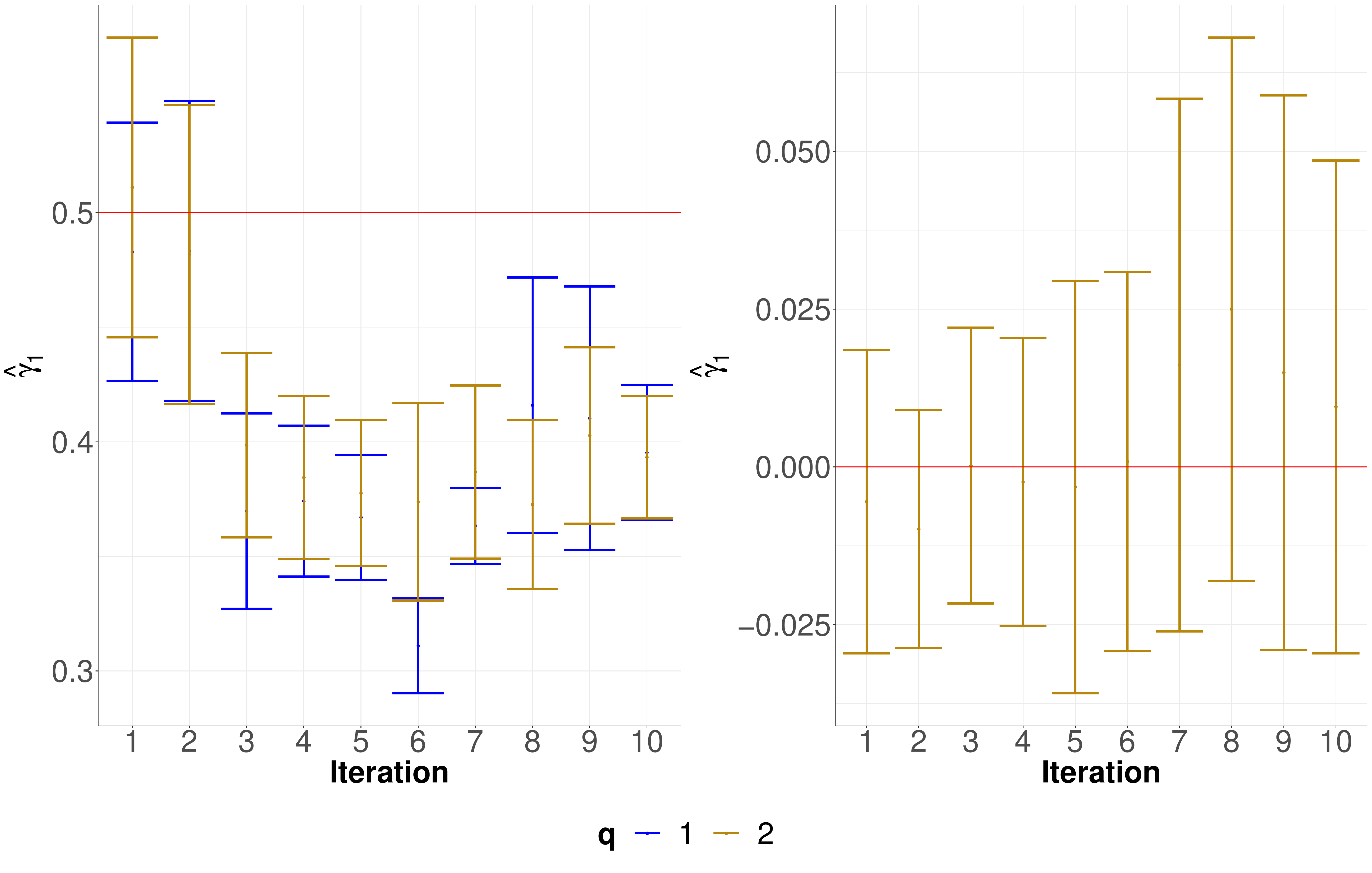}
\caption{Error bars  for the estimations of $\boldsymbol{\gamma}^\star$ in Model \eqref{eq2} for  $I=3$, $T=50$, $J =10$, $q^{\star}$ = 1, $\gamma^{\star} = 0.5$,
  10 non-null coefficients in $\pmb{\eta^{\star}}$, and 50 simulations obtained by $q=1$. The horizontal lines correspond to the values of the $\gamma^{\star}_i$’s.}
\end{figure}

\begin{figure}[!htb]
\centering
\includegraphics[scale=0.25]{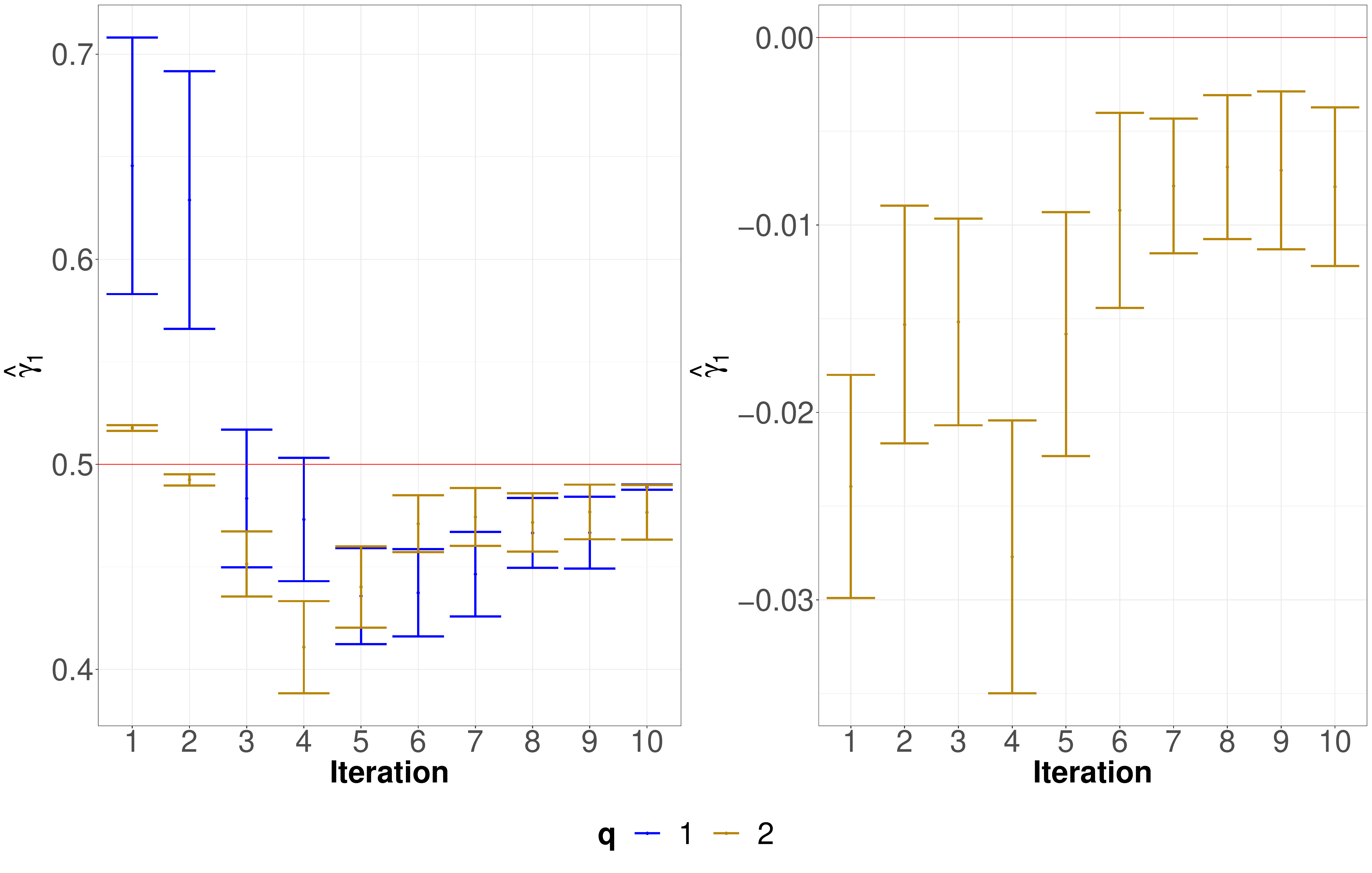}
\caption{Error bars for the estimations of $\boldsymbol{\gamma}^\star$ in Model \eqref{eq2} for  $I=3$, $T=50$, $J =100$, $q^{\star}$ = 1, $\gamma^{\star} = 0.5$,
  10 non-null coefficients in $\pmb{\eta^{\star}}$, and 50 simulations obtained by $q=1$. The horizontal lines correspond to the values of the $\gamma^{\star}_i$’s.}
\end{figure}
\FloatBarrier

\newpage
\subsection{Additional results for the application section}
\label{app:app}

\vspace{30mm}

{\scriptsize{
\begin{table}[H]
\begin{tabular}{|c|c|c|c|c|c|c|}
\hline
\backslashbox{Gene}{Temp.pop}    & Low.npoly                 & Medium.npoly                 & Elevated.npoly               & Low.poly            & Medium.poly            & Elevated.poly            \\ \hline
AT1G28230 & 240                & 52.33              & 21.33 & 268.67   & 56                 & 25.33   \\ \hline
AT1G18100 & 696                & 162.33   & 98.67 & 1385.33 & 249                & 219                \\ \hline
AT1G48130 & 6151.67  & 1179.33 & 841              & 6170.67  & 1511.33 & 1162.33 \\ \hline
AT1G47980 & 2947               & 346.67  & 176.33 & 2114.33 & 418.33  & 198                \\ \hline
AT1G14940 & 2277.67 & 689.67  & 598              & 3661               & 776.67  & 832                \\ \hline
AT1G14950 & 4630.67  & 1180.33 & 476              & 6617               & 1465               & 883                \\ \hline
AT1G07090 & 320.33  & 189                & 81.67 & 557.33  & 286.33  & 174.33   \\ \hline
AT1G70810 & 499                & 221                & 137.33 & 1161.33 & 410.67  & 305                \\ \hline
AT1G73190 & 2860.33 & 502                & 236.33 & 3090               & 608.33  & 335.33  \\ \hline
AT2G33830 & 2498.33 & 379.33  & 305              & 7977.33  & 1307               & 1110               \\ \hline
AT2G26010 & 50.67   & 7.67   & 1.33 & 150.67   & 17.33   & 3.67   \\ \hline
AT2G43510 & 1350.33 & 305.33  & 162.33 & 3943.67 & 887.67  & 514                \\ \hline
AT2G29300 & 248.33   & 70.33   & 32.33 & 253.67   & 62.33   & 34                 \\ \hline
AT2G43580 & 664                & 91                 & 31.33 & 694.33  & 88                 & 37.33   \\ \hline
AT2G23640 & 285.67  & 140                & 122.33 & 830                & 262                & 328                \\ \hline
AT3G56350 & 2105.67 & 442                & 223.67 & 2785.67 & 542      & 345                \\ \hline
AT4G24715 & 1097.33 & 263.33  & 116.67 & 1249.67 & 318.67  & 202                \\ \hline
AT4G02520 & 3852               & 866.67  & 621              & 2575.33 & 656.33  & 450.67  \\ \hline
AT4G00220 & 180.33   & 98.33   & 66               & 367.33 & 122.33   & 108.33   \\ \hline
AT4G26740 & 2026               & 442.67  & 266              & 2606.67 & 526                & 393.67  \\ \hline
AT5G55560 & 297.67   & 138                & 72               & 603                & 266.67   & 158.67   \\ \hline
\end{tabular}
\caption{Data used for displaying Figure \ref{fig:application} at the different temperature (Low, Medium, ELevated)
  where ``.npoly'' (resp. ``.poly'') are the values corresponding to the non-polysomal (resp. polysomal) population.}
\label{tab_ap_1}
\end{table}
}}
\FloatBarrier

\begin{figure}[!htb]
\centering
\includegraphics[scale=0.25]{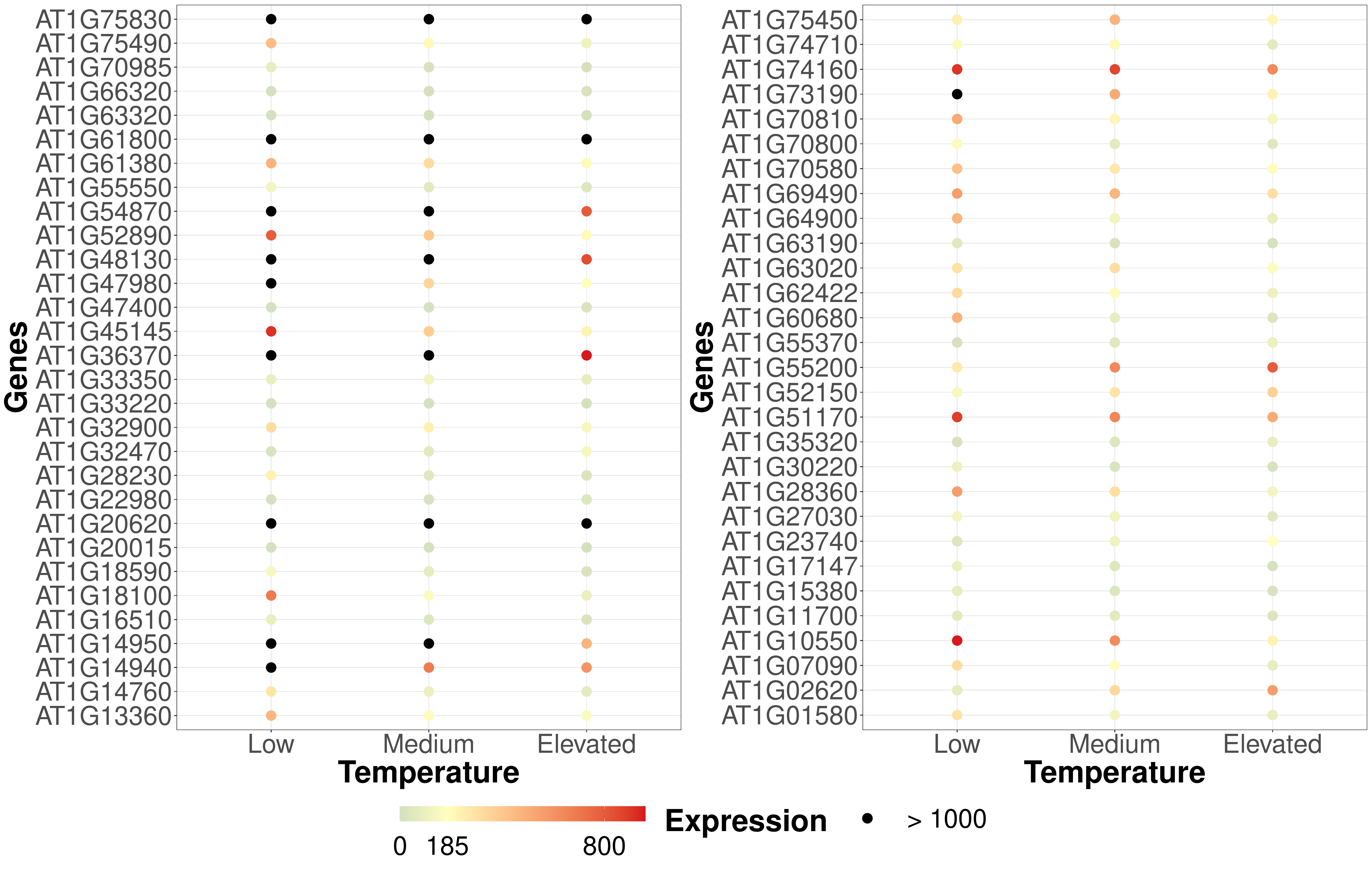}
\caption{Average gene expression values of 3 replications at each temperature condition
  (Low, Medium, Elevated) for the chromosome 1 in non-polysomal population. The genes displayed in this figure are obtained by our selection procedure.}
\label{fig:non_poly_1}
\end{figure}
\FloatBarrier

\begin{figure}[!htb]
\centering
\includegraphics[scale=0.25]{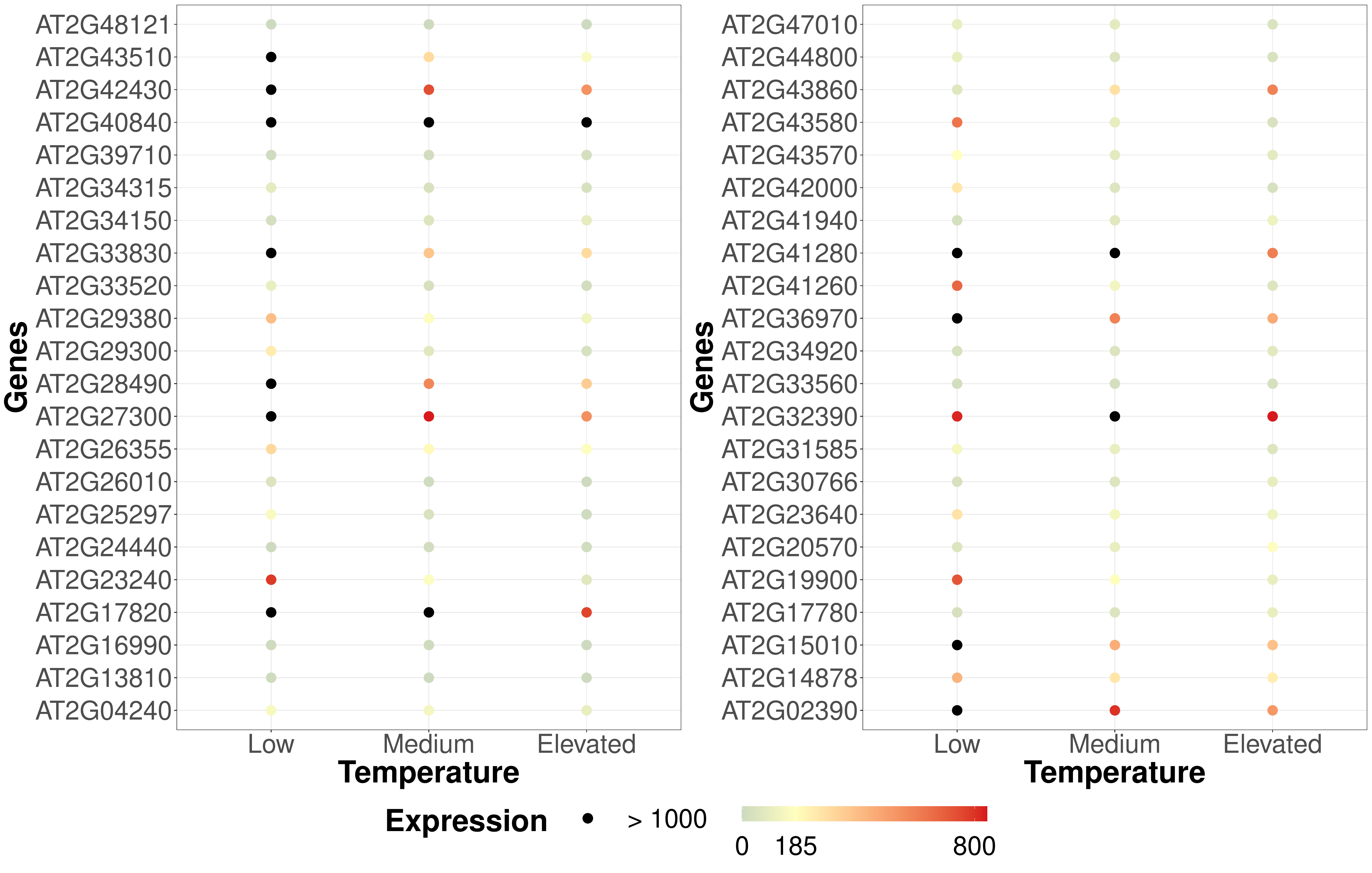}
\caption{Average gene expression values of 3 replications at each temperature condition
  (Low, Medium, Elevated) for the chromosome 2 in non-polysomal population. The genes displayed in this figure are obtained by our selection procedure.}
\label{fig:non_poly_2}
\end{figure}
\FloatBarrier

\begin{figure}[!htb]
\centering
\includegraphics[scale=0.25]{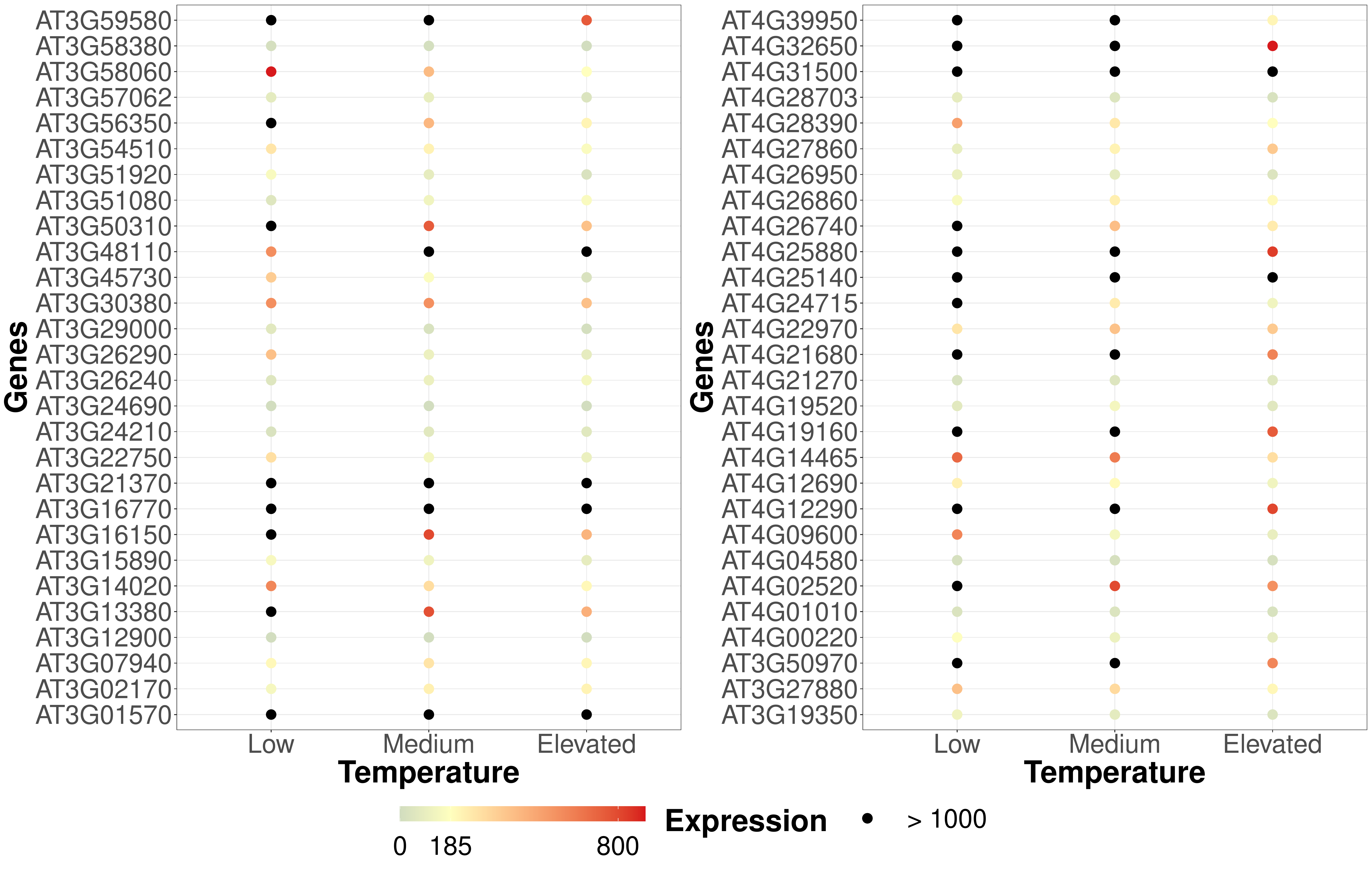}
\caption{Average gene expression values of 3 replications at each temperature condition
  (Low, Medium, Elevated) for the chromosomes 3 and 4 in non-polysomal population. The genes displayed in this figure are obtained by our selection procedure.}
\label{fig:non_poly_3_4}
\end{figure}
\FloatBarrier

\begin{figure}[!htb]
\centering
\includegraphics[scale=0.25]{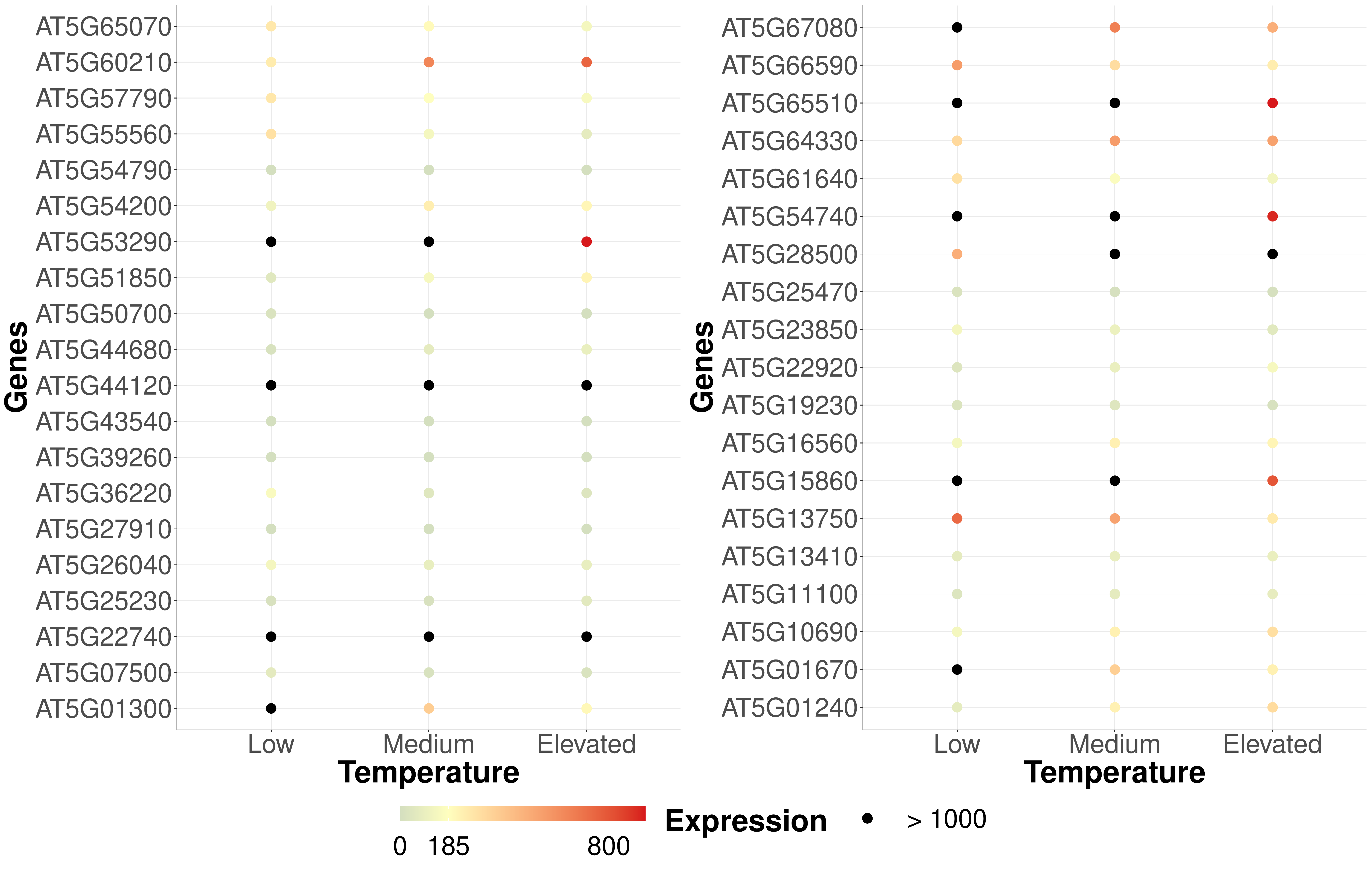}
\caption{Average gene expression values of 3 replications at each temperature condition
  (Low, Medium, Elevated) for the chromosome 5 in non-polysomal population. The genes displayed in this figure are obtained by our selection procedure.}
\label{fig:non_poly_5}
\end{figure}
\FloatBarrier

\begin{figure}[!htb]
\centering
\includegraphics[scale=0.25]{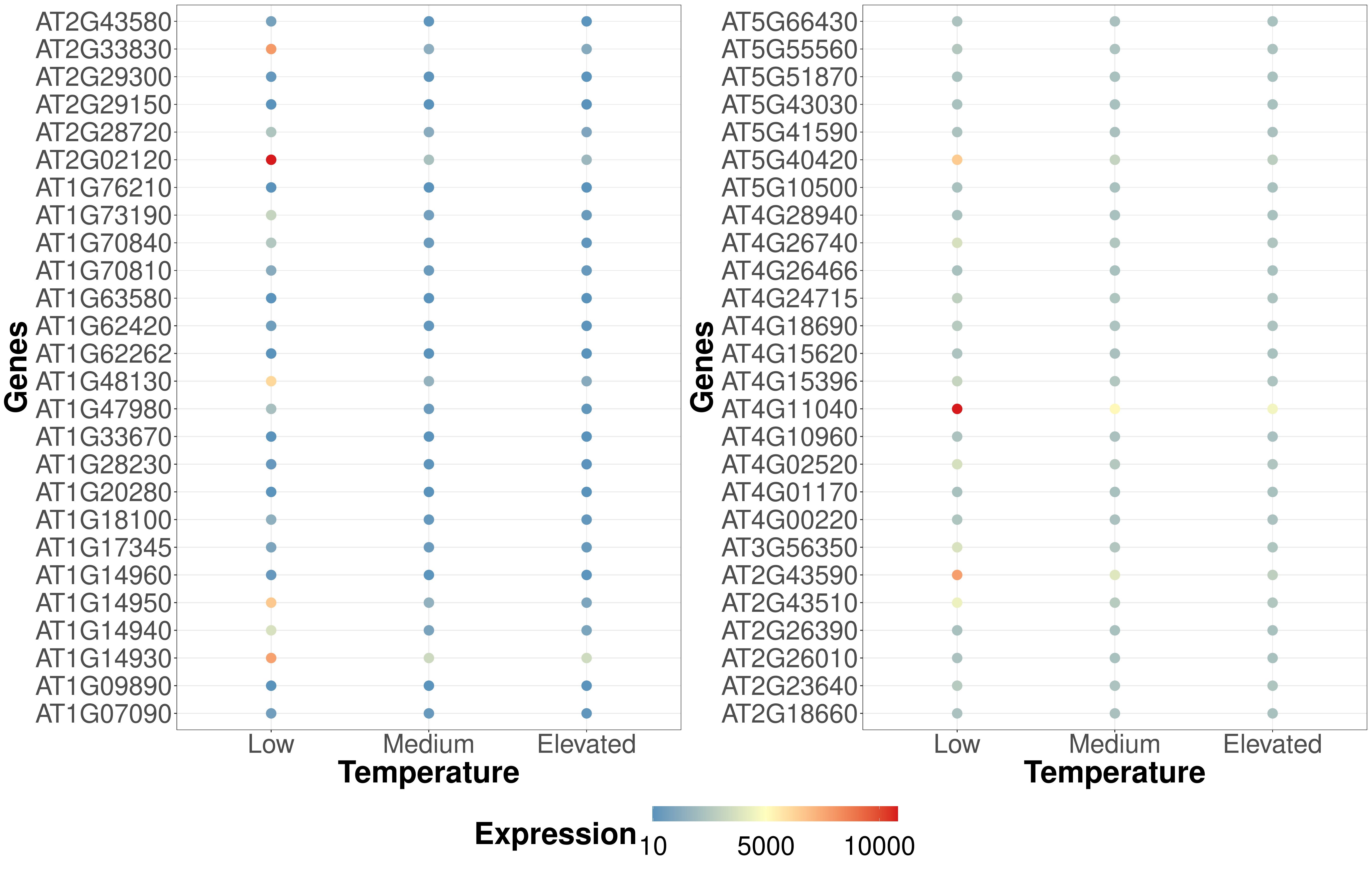}
\caption{Average gene expression values of 3 replications at each temperature condition
  (Low, Medium, Elevated) for the polysomal population. The genes displayed in this figure are obtained by our selection procedure.}
\label{fig:poly}
\end{figure}
\FloatBarrier

\bibliographystyle{chicago}
\bibliography{biblio}
\end{document}